\documentclass[superscriptaddress,amsmath,twocolumn,amssymb,prd,preprintnumbers,showpacs]{revtex4-2}
\usepackage{amssymb}
\usepackage{amsopn}
\usepackage{hyperref}
\usepackage{xcolor}
\usepackage[caption=false]{subfig}
\usepackage{soul}
\usepackage{amsmath}
\usepackage{graphicx}
\usepackage{float}
\usepackage{amsopn}
\usepackage{braket}
\usepackage{slashed}
\usepackage{bbold}
\newcommand{\sla}[1]{\hbox{{$#1$}\llap{$/$}}}

\begin{document}
%.................................................................
\title{Tree-level unitarity, causality and higher-order Lorentz and CPT violation}
\author{Justo L\'opez-Sarri\'on}
\email[Electronic mail:]{ justo.lopezsarrion@ub.edu}
\affiliation{\it Departament de F\'\i sica Qu\`antica i Astrof\'\i sica and \\
Institut de Ci\`encies del Cosmos (ICCUB), Universitat de Barcelona,\\ 
Mart\'i  Franqu\`es 1, 08028 Barcelona, Spain}
\author{Carlos M. Reyes}
\email[Electronic mail: ]{ creyes@ubiobio.cl}
\affiliation{ Centro de Ciencias Exactas, Universidad del B\'{i}o-B\'{i}o \\
Avda.~Andr\'es Bello 720, Chill\'{a}n, 3800708, Chile }
\author{ C\'esar Riquelme}
\email[Electronic mail:]{ ceriquelme@udec.cl}
\affiliation{ Centro de Ciencias Exactas, Universidad del B\'{i}o-B\'{i}o \\
Avda.~Andr\'es Bello 720, Chill\'{a}n, 3800708, Chile }
\affiliation{ Departamento de F\' {\i}sica, Universidad de Concepci\'on, Casilla 160-C, Concepci\'on, Chile }
%...............................................................................................................................................
\begin{abstract}
%...............................................................................................................................................
Higher-order effects of CPT and Lorentz violation within the SME effective framework 
including Myers-Pospelov dimension-five operator terms are studied. 
The model is canonically quantized by giving 
special attention to the arising of indefinite-metric states or ghosts in an indefinite Fock space. 
As is well-known, without a perturbative treatment that avoids the 
propagation of ghost modes or any other approximation, one 
has to face the question of whether unitarity and microcausality 
are preserved. In this work, we study both possible issues. 
We found that microcausality is preserved due to the cancellation 
of residues occurring 
in pairs or conjugate pairs when they become complex. Also, by 
using the Lee-Wick prescription,
we prove that the $S$ matrix can be defined as perturbatively unitary 
for tree-level $2\to 2$ processes with an internal fermion line.
\end{abstract}
%...........................................................................
\maketitle
%...........................................................................
\section{Introduction}
%...........................................................................
Quantum field theory (QFT) is conceptually based
 on locality and Lorentz invariance. 
Any departure from these two basic concepts will introduce serious alterations to the traditional construction of 
field theory and will necessarily imply new physics. Alternative theories containing Lorentz invariance violation have been widely studied
to test the limits of conventional QFT. 
The triad of theoretical, phenomenological, and experimental work has made significant progress in the last two decades.
 In particular, the search for potential Lorentz violations has received special
  attention producing stringent limits on Lorentz violations with ultrahigh sensitive experiments~\cite{review,tables}.

The fundamental interplay between matter and geometry 
continues to be a source of conceptual issues. At the Planck mass $m_{\text{Pl}}\approx 10^{19}$ 
GeV, various candidate theories of quantum gravity suggest the disruption of the continuum property of 
spacetime. If Minkowski spacetime is not the exact geometry at these energies, then it is justified to
 consider the standard model of particles to be an effective theory. One should expect experiments 
 taking place at scales $\Lambda$ to describe gravitational effects suppressed by
 $\Lambda/m_{\text{Pl}}$. Nevertheless, residual gravitational effects could be detected at currently 
 attainable energies. A possible manifestation of such disruption has been realized in the form of CPT and
Lorentz violations~\cite{strings,strings2,LQG}. In this way, the search for possible effects of Lorentz violation 
using effective field theory has been amply 
adopted. Effective field theory has become a natural language in high-energy phenomenology to 
describe possible Lorentz violations.
This work focuses on the possible effects of CPT and Lorentz violation described within 
an effective framework.

The effective framework of the Standard-Model Extension (SME) 
describes effects of CPT and Lorentz violation in field theory by introducing gauge-invariant objects 
constructed from Standard-Model fields coupled to vectors and tensors that parametrize the Lorentz 
violation. It also covers the gravity sector where local Lorentz and diffeomorphism violation give rise to modified-gravity
theories. The SME can be divided into a minimal sector and a nonminimal sector. 
The minimal sector includes renormalizable operators of mass dimensions equal or lower than four, and it was the 
first sector to be proposed~\cite{smext}. The natural next step was to focus on higher-order operators 
with mass dimensions five or higher, which has been carried out extensively in the past years, giving several bounds 
on the parameters that modify QFT~\cite{HOphotons, HOfermions} and linearized gravity~\cite{HOgravity}. The 
Myers-Pospelov model was formulated independently and focused on dimension-five operators containing 
Lorentz violation in the scalar, fermion, and photon sectors~\cite{MP,P2}.
Consistency properties such as causality, stability~\cite{Causality_Sability0,Causality_0,Causality_Sability,Causality_Sability2} 
and unitarity in the minimal~\cite{Unitarity0,Unitarity1} and nonminimal sectors of the SME~\cite{Unitarity2,Unitarity3,CMR,C-odd}
have been studied intensively in the past years. 
Also, theories of fermions and photons with broken spin degeneracy have been studied in~\cite{Schreck_fermions-photons2}. 
This class of theories provides the possibility to open a window to effects relying on a nonzero phase space, such 
as Cherenkov radiation in vacuo and decay of photons into electron-positron pairs~\cite{Klinkhamer,MSV1}. 
Radiative corrections have also been extensively studied within the SME~\cite{rad}.
Recently a sector of modified gravity has been cast in canonical form~\cite{ADM12}, and Lorentz-violating cosmology 
has been proposed~\cite{LIVcosmology}.

The effects introduced by higher-order operators become 
stronger at higher energies since they scale with higher powers of momenta. 
However, a notable nonperturbative effect is that 
they generically introduce extra degrees of freedom associated with negative-norm states in an indefinite 
Hilbert space. Contrary to the Gupta-Bleuler formalism in covariant QED~\cite{GB} 
the negative-norm states associated with higher-order operators
 can not be a priori excluded from the asymptotic state space. A treatment introduced 
 by Lee and Wick in which a specific asymptotic space is adopted
 successfully proved that theories with indefinite metric can preserve 
 unitary, thereby respecting the probability interpretation of quantum mechanics~\cite{LW,Boulware_Gross}.
Indefinite Hilbert spaces may lead to the loss of unitarity. The 
negative-metric part associated with ghost states can modify the 
amplitudes, disrupting the optical theorem, being a direct consequence of unitarity. 
In this work, we investigate the preservation of unitarity in a process
of QED involving $2\to2$ particles at tree-level. 
We have focused on the extension  
of the Myers and Pospelov fermion sector that is even under charge conjugation (C).
In particular, 
the C-odd part has been studied in~\cite{C-odd}.

The organization of this work is as follows. In Sec.~\ref{sectionII} we compute the 
dispersion relations and find the spinor solutions. In Sec.~\ref{sectionIII} we 
quantize the fermion sector, find the Hamiltonian and compute the propagator 
using its definition in terms of expectation values of the fields. 
Furthermore, in Sec.~\ref{sectionIV} we compute the Pauli-Wigner function 
for two separated spacetime points and verify microcausality. In Sec.~\ref{sectionV} 
we compute unitarity at tree-level in $2\to2$ particles processes by using the optical 
theorem. Section~\ref{sectionVI} contains our final remarks.
%.....................................................................................................
\section{Higher-order Lorentz violating model  }\label{sectionII}
%....................................................................................................
We start with the higher-order Lorentz and CPT-violating 
Lagrangian proposed in~\cite{MP}
\begin{align} \label{Mod_fermions}
\mathcal{L}_F=\bar \psi (i \slashed{\partial}-m ) \psi  
+\frac{  n^{\mu} n^{\nu}   } {m_{\textrm{Pl}} }\bar
 \psi (\eta_1 \slashed{n}+\eta_2 \slashed{n} \gamma_5)
    (\partial_{\mu}\partial_{\nu})  \psi  \,,
\end{align}
where $n^\mu$ is a constant four-vector, $\eta_1$ and $\eta_2$ are constants couplings
being charge conjugation odd and even, respectively. As usual
$m_{\textrm{Pl}}$ is the Planck mass.

The free equation of motion is 
\begin{equation} \label{free_motion}
\left( i \slashed{\partial}-m   +\frac{  n^{\mu} n^{\nu}   } {m_{\textrm{Pl}} } 
(\eta_1 \slashed{n}+\eta_2 \slashed{n} \gamma_5)
    (\partial_{\mu}\partial_{\nu}) \right) \psi(x) =0 \,.
\end{equation}
The gauge-invariant
QED Lagrangian can
be obtained via minimal 
coupling substitution in~\eqref{Mod_fermions}, producing 
\begin{align} \label{TotalLagrangian}
\mathcal{L}_{\text{QED}}&=\bar \psi (i \slashed{D}-m ) \psi  +\frac{n^{\mu} n^{\nu}}{m_{\textrm{Pl}}}\bar
 \psi (\eta_1 \slashed{n}    +\eta_2 \slashed{n} \gamma_5)
  \nonumber \\ &\times    D_{\mu}D_{\nu}   \psi -\frac{1}{4} F_{\mu\nu}F^{\mu\nu} \,, 
\end{align}
where $D_\mu=\partial_\mu+ieA_\mu$ and $F_{\mu \nu}=\partial_{\mu} A_{\nu}-\partial_{\nu} A_{\mu}$. 

Consider the gauge transformations on the fields
 \begin{eqnarray}
A_{\mu} (x)&\to& A_{\mu}(x)+ \partial_{\mu}\lambda(x)\,, \nonumber  \\ \label{GT}
\psi(x) &\to& e^{-ie\lambda}\psi (x)\,,
 \end{eqnarray}
one can prove they lead to
\begin{eqnarray}
D_{\mu} \psi &\to& e^{-i e \lambda}D_{\mu}  \psi \,.
\end{eqnarray}
 Thus, the gauge invariance of the Lagrangian~\eqref{TotalLagrangian}
 follows from the transformation
 \begin{eqnarray}
D_{\alpha} ( e^{-ie\lambda} D_{\mu} \psi )&\to& \partial_{\alpha}  (e^{-ie\lambda} 
D_{\mu} \psi )  +ie (A_{\alpha}+  \partial_{\alpha}\lambda) \nonumber
 \\ &\times& e^{-ie\lambda} D_{\mu} \psi  \nonumber \\
&=& e^{-i e \lambda} D_{\alpha}D_{\mu}  \psi  \,.
\end{eqnarray}
Here we 
work with the Dirac matrices in the chiral representation, i.e,
\begin{equation}
\gamma^\mu =\left( \begin{array}{c c}
0\quad & \sigma^\mu\\
\bar\sigma^\mu  \quad &0
\end{array}\right)\,, \qquad 
\gamma_5=\left(\begin{array}{c c}
-\mathbb{1}_{2} \quad &0\\
0 \quad &\mathbb{1}_{2} \end{array}\right)\,,
\end{equation}
where 
$\sigma^\mu=(\mathbb{1}_{2},\vec\sigma)$,
$\bar\sigma^\mu=(\mathbb{1}_{2},-\vec\sigma)$ and 
$\mathbb{1}_{2}$ is the $2\times2$ identity matrix.
The fields are defined in Minkowski spacetime with 
metric signature $(+,-,-,-)$.
%.....................................................................................................
\subsection{The dispersion relation}
%....................................................................................................
For the rest of the work we turn off the charge conjugation odd 
sector setting $\eta_1=0$ in the Lagrangian~\eqref{Mod_fermions}.

Consider the ansatz $\psi(\vec x)=\int d^3\vec p \; u( p) e^{-ip\cdot x}$ substituted in
Eq.~\eqref{free_motion}. We arrive at
\begin{equation}\label{equation_spinor}
    \left(\sla{p}-m-g_2\slashed{n}\gamma_5(n\cdot p)^2\right) u( p)=0\,,
\end{equation}
with the redefined coupling $g_2\equiv \eta_2/m_{Pl}$.

Let us define the operators
\begin{align}
	\mathcal M&=\slashed{p}-m-g_2\slashed{n}\gamma_5(n\cdot p)^2 \notag  \,, \\
	\bar { \mathcal M}&=\slashed{p}+m-g_2\slashed{n}\gamma_5(n\cdot p)^2\,,
\end{align}	
and
\begin{align}	
	{\mathcal N}&=\slashed{p}+m+g_2\slashed{n}\gamma_5(n\cdot p)^2 
	\notag \,,\\ \bar{\mathcal N}&=\slashed{p}-m+g_2\slashed{n}\gamma_5(n\cdot p)^2 \label{barN} \,.
\end{align}
In addition we define 
\begin{align}\label{Q}
{	\mathcal Q}&=-	\frac{ \left[ \slashed{p},\slashed{n} \right] \gamma_5}{2 \sqrt{D}}  \,,
\end{align}
where $D(n,p):= (n\cdot p)^2-p^2n^2 $ is the Gramian of
 the two four-vectors $n$ and $p$. The operator $ { \mathcal Q}$, 
commutes with the equation of motion, i.e.,
\begin{eqnarray}
[{\mathcal Q}, {\mathcal M}]=0   \,,
\end{eqnarray}
and with any of the operators $\bar {\mathcal M}, \mathcal N,\bar {\mathcal N}$, so we expect 
the spinor solutions to be eigenstates of $\mathcal Q$.

Some useful relations follows by considering 
\begin{align}
	\bar {\mathcal M}  \mathcal M &=p^2-m^2-g_2^2n^2 (n\cdot p)^4
	\nonumber  \\ &+2g_2(n\cdot p)^2  \sqrt{D}\,  {\mathcal Q} \,,
\end{align}
and
\begin{align}
	\bar {\mathcal N} \mathcal N  &=p^2-m^2-g_2^2n^2 (n\cdot p)^4 \nonumber  
	\\ &-2g_2(n\cdot p)^2   \sqrt{D} \, {\mathcal Q} \,.
\end{align}
We have
\begin{align}\label{eq_decomp}
\left( \bar {\mathcal N} \mathcal N\bar {\mathcal M}
 \mathcal M \right)u(p) &=\left(\left (p^2-m^2-g_2^2n^2 (n\cdot p)^4 \right)^2
 \nonumber  \right. \\  &- \left. 4g_2^2 (n\cdot p)^4 D \right)u(p)=0\,,
\end{align}
where it has been used the identities
\begin{eqnarray}
	\left[ \slashed{p},\slashed{n} \right] \gamma_5
	 \left[ \slashed{p},\slashed{n} \right] \gamma_5&=&4 D  \,,
\end{eqnarray}
and
\begin{eqnarray}
 {	\mathcal Q}^2&=&1  \,.
\end{eqnarray}
We arrive at the dispersion relation by requiring a nontrivial solution for $ u(p)$, that is to say
\begin{eqnarray}\label{Disp_Rel}
   \left (p^2-m^2-g_2^2n^2 (n\cdot p)^4 \right)^2-4g_2^2 (n\cdot p)^4 D =0\,.
\end{eqnarray}
Let us define the two quantities
\begin{eqnarray} \label{disp_pieces1}
	\widetilde \Lambda_+^2(p)&=&p^2-m^2-g_2^2n^2 (n\cdot p)^4 
	\nonumber \\&-&2g_2 (n\cdot p)^2 \sqrt{ D} \,,
\end{eqnarray}	
and	
\begin{eqnarray} 	\label{disp_pieces2}
	\widetilde \Lambda_-^2(p)&=& p^2-m^2-g_2^2n^2 (n\cdot p)^4 
	\nonumber \\&+&2g_2 (n\cdot p)^2 \sqrt{ D} \,.
\end{eqnarray}
Their product produce the dispersion relation
\begin{eqnarray}\label{dispersion_purelytimelike}
	\widetilde \Lambda_+^2(p) \widetilde \Lambda_-^2(p)&\equiv &
	\left (p^2-m^2-g_2^2n^2 (n\cdot p)^4 \right)^2 \nonumber \\   &-&4g_2^2 (n\cdot p)^4 D \,.
\end{eqnarray}
%...................................................................................
\subsection{Purely timelike model}
%................................................................................... 
Here we consider the background to be purely timelike with $n=(1,0,0,0)$.
Hence, the Lagrangian~\eqref{Mod_fermions} takes the form
\begin{equation}\label{HO_Lag}
\mathcal{L}=\bar \psi (i \slashed{\partial}-m ) \psi  +g_2\bar \psi \gamma_0 \gamma_5  \ddot{\psi}\,,
\end{equation}
with equation of motion 
 \begin{equation}\label{timelike_spinor}
    \left(\sla{p}-m-g_2 p_0^2 \gamma_0 \gamma_5  \right) \psi( p)=0\,.
\end{equation}
 The previous operators are now
 \begin{eqnarray}
	 M&=&\slashed{p}-m-g_2p_0^2  \gamma_0  \gamma_5 \,,  \label{Op_M}\\
	\bar {  M}&=&\slashed{p}+m-g_2 p_0^2 \gamma_0\gamma_5  \label{Op_Mbar} \,,\\
	 { N}&=&\slashed{p}+m+g_2 p_0^2 \gamma_0\gamma_5  \,, \label{Op_N}\\ 
  \bar {N}&=&\slashed{p}-m+g_2p_0^2 \gamma_0\gamma_5 \,.
\end{eqnarray}
Furthermore, we have 
\begin{equation}\label{Qtime}
	 Q =-\frac{p_i\gamma^i}{|\vec p|}   \gamma_0 \gamma_5 =- \left( \begin{array}{c c}
  \frac{ \vec \sigma \cdot \vec p}{|\vec p|} \quad & 0  \\
0 \quad & \frac{ \vec \sigma \cdot \vec p}{|\vec p|}
\end{array}\right)  \,,
\end{equation}
and
\begin{eqnarray}\label{lambdas}
 \Lambda_{+}^2 (p)&=& p_0^2-|\vec{p}|^2 -m^2 -   
g_2^2 p_0^4-2 g_2p_0^2 |\vec{p}|\,,
\nonumber \\
 \Lambda_{-}^2(p) &=& p_0^2-|\vec{p}|^2 -m^2 - g_2^2 p_0^4+2 g_2p_0^2|\vec{p}|\,,
\end{eqnarray}
which can be rewritten as 
\begin{eqnarray}\label{DRpart}
		 \Lambda_{+}	^2 +m^2&=& (p_0 + g_2 p_0^2 + |\vec{p}|) (p_0 - g_2 p_0^2 - |\vec{p}|)  \,,
\nonumber \\
		 \Lambda_{-}	^2 +m^2&=&  (p_0 + g_2 p_0^2 - |\vec{p}|) (p_0 - g_2 p_0^2 + |\vec{p}|)   \,.
\end{eqnarray}
The dispersion relation Eq.~\eqref{dispersion_purelytimelike} is
\begin{eqnarray}\label{SimpDispersion}
	(p_0^2-  |\vec{p}|^2-m^2-g_2^2 p_0^4)^2-4g_2^2 p_0^4|\vec{p}|^2=0\,.
\end{eqnarray}

The eight solutions to the dispersion relations come from two sectors. 
We have four solutions of the dispersion relation $ \Lambda_+^2=0$ 
\begin{eqnarray} \label{posit_freq}
	\omega_1&=&\sqrt{\frac{1-2g_2 |\vec p |- \sqrt{(1-2g_2 |\vec p |  )^2 -
	4g_2^2E_p^2}}{2g_2^2} }\,, \nonumber 
\\ \overline \omega_1&=&-\omega_1\,,	\nonumber  \\
	W_1&=&\sqrt{\frac{1-2g_2 |\vec p | + \sqrt{(1-2g_2 |\vec p |  )^2 -
	4g_2^2E_p^2}}{2g_2^2} }\,, \nonumber  \\ \overline W_1&=&-W_1\,,
\end{eqnarray}
and four solutions of the dispersion relation $ \Lambda_-^2=0$ 
\begin{eqnarray}  \label{negat_freq}
	\omega_2&=&\sqrt{\frac{1+2g_2 |\vec p |- \sqrt{(1+2g_2 |\vec p |  )^2
	 -4g_2^2E_p^2}}{2g_2^2} }\,, \nonumber 
	\\  \overline \omega_2&=&-	\omega_2\,,  \nonumber    \\
	W_2&=&\sqrt{\frac{1+2g_2 |\vec p | + \sqrt{(1+2g_2 |\vec p |  )^2 
	-4g_2^2E_p^2}}{2g_2^2} } \,,\nonumber  \\ \overline W_2&=&-W_2\,,
\end{eqnarray}
where $E_p=\sqrt{ |\vec p |^2+ m^2}$.

Alternatively, we can rewrite the total dispersion relation as
\begin{eqnarray}
		 \Lambda_+^2(p)	 \Lambda_-^2(p) &=&  g_2^4  (p_0^2-\omega_1^2)    (p_0^2-W_1^2)
		  (p_0^2-\omega_2^2) \nonumber \\ &\times&  (p_0^2-W_2^2)=0\,.
\end{eqnarray}

The solutions can be analyzed individually, let us expand for small coupling, and obtain up to
linear order in $g_2$
\begin{eqnarray}
\omega_1&\approx&E_p +   |\vec p | E_p  g_2\,,
\\
\omega_2&\approx&E_p -    |\vec p | E_p  g_2\,,
\\
  W_1&\approx& \frac{1}{g_2}-  |\vec p | -\frac{1}{2}(E^2_p+  |\vec p |^2)g_2  \,,  
\\
W_2&\approx& \frac{1}{g_2}+  |\vec p | -\frac{1}{2}(E^2_p.+  |\vec p |^2)g_2    \,.
\end{eqnarray}
The low-energy modes $\omega _1$ and $\omega _2$ are perturbatively connected to particle propagation, 
however, the additional degrees of freedom corresponding the the higher-energy modes $W _1$ and $W _2$
correspond to the propagation of negative-norm states or ghosts as we will show in the next sections.

The frequencies $\omega_1, W_1$ and $\overline \omega_1, \overline W_1$
 can become complex for higher momenta. The condition for this to occur is
\begin{align}
 (1-2g_2 |\vec p |  )^2 - 4g_2^2E_p^2<0\,,
\end{align}
 from where we find  a region where energies become complex $\vert p\vert >\vert p_{\max}\vert=\frac{1-4g_2^2m^2}{g_2}$. Note that 
 the condition for energies $\omega_2, W_2$ and  $\overline \omega_2, \overline W_2$
 \begin{align}
  (1+2g_2 |\vec p |  )^2 - 4g_2^2E_p^2<0\,,
 \end{align}
 can not be satisfied for small values of $g_2^2m^2$ and hence the energy
  remain real for any momenta.
 We find
  \begin{align}
  \omega_1( \vert p_{\max}\vert  )= W_1(\vert p_{\max}\vert)=\frac{1}{2}\sqrt{\frac{1}{g_2^2}+4m^2}  \,,
  \end{align}
 and $ \lim_{\vert p\vert\rightarrow \infty} \omega_{2}= \lim_{\vert p\vert\rightarrow \infty} W_{2}\rightarrow \infty $.
At this level, the theory establishes a maximum value for the momentum and a priori 
an energy scale for the effective region of the theory. 
%...................................................................................
\subsection{Spinor solutions}
%...................................................................................
Now we focus on finding the eigenspinors 
of the modified Dirac equation using the energy solutions~\eqref{posit_freq} and \eqref{negat_freq}.
Consider the field $\psi(\vec x)=\int d^3\vec p \; u( p) \; e^{-ip\cdot x}$ in the equation of motion~\eqref{timelike_spinor}
which produces
\begin{eqnarray}\label{eqM}
	M u(p)=0 \,,
\end{eqnarray}
where $M$ defined in Eq.~\eqref{Op_M} has matrix form
\begin{align}\label{eq_motion}
	 M=& \left(\begin{array}{c c}
		-m& p_0-g_2p_0^2-(\vec{p}\cdot\vec{\sigma} )   \\
		p_0+g_2p_0^2+(\vec{p}\cdot\vec{\sigma} ) &-m
	\end{array}\right)  \,.
\end{align}
We write the spinor in terms of bi-spinors
\begin{equation}\label{Positivefrequency}
u( p)=
\left(\begin{array}{c} \chi_1(p)\\ \chi_2(p)\end{array}\right) \,,
\end{equation}
and arrive at the equations
\begin{eqnarray}
(p_0-g_2p_0^2-(\vec{p}\cdot\vec{\sigma} ))   \chi_2=m\chi_1\,, \nonumber \\ \label{posit_sol}
(	p_0+g_2p_0^2+(\vec{p}\cdot\vec{\sigma} ) )  \chi_1=m\chi_2\,.
\end{eqnarray}
The spinor solutions of the dispersion relation $ \Lambda^2_+=0$
are
\begin{eqnarray}
	u^{(1)}( p)&=&\left(\begin{array}{c}   \sqrt{p_0-g_2p_0^2-|\vec{p}| } \xi^{(+)}(\vec p)\\ 
	 \sqrt{p_0+g_2p_0^2+|\vec{p}| }  \xi^{(+)}(\vec p) \end{array}\right)_{p_0=\omega_1}  \,,
\nonumber \\
	U^{(1)}( p)&=&\left(\begin{array}{c}  \sqrt{p_0-g_2p_0^2-|\vec{p}| }\xi^{(+)}(\vec p)\\ \label{spinors1}
	\sqrt{p_0+g_2p_0^2+|\vec{p}| }\xi^{(+)}(\vec p) \end{array}\right)_{p_0=W_1}  \,.
\end{eqnarray}
and the solutions of the dispersion relation $ \Lambda^2_-=0$
\begin{eqnarray}
	u^{(2)}( p)&=&\left(\begin{array}{c} \sqrt{p_0-g_2p_0^2+|\vec{p} |}  \xi^{(-)}(-\vec p)\\ 
	\sqrt{p_0+g_2p_0^2-|\vec{p} |}  \xi^{(-)}(-\vec p) \end{array}\right)_{p_0=\omega_2}  \,,
\nonumber \\
	U^{(2)}( p)&=&\left(\begin{array}{c}  \sqrt{p_0-g_2p_0^2+|\vec{p} |}  \xi^{(-)}(-\vec p)\\ \label{spinors2}
	\sqrt{p_0+g_2p_0^2-|\vec{p} |} \xi^{(-)}(-\vec p) \end{array}\right)_{p_0=W_2}  \,.
\end{eqnarray}
For the negative-energy solutions, we consider the field to be 
$\psi(\vec x)=\int d^3\vec p \; v( p) \; e^{ip\cdot x}$
and the 
eigenvalue equation 
\begin{eqnarray}\label{neg_sol}
	N v(p)=0 \,,
\end{eqnarray}
with
\begin{eqnarray}\label{eq_motionN}
	 	N= \left(\begin{array}{c c}
		m& p_0+g_2p_0^2-(\vec{p}\cdot\vec{\sigma} )   \\
		p_0-g_2p_0^2+(\vec{p}\cdot\vec{\sigma} ) &m
	\end{array}\right)  \,, \nonumber \\
\end{eqnarray}
given in Eq.\eqref{Op_N}
and 
\begin{equation}\label{Positivefrequency2}
v( p)=
\left(\begin{array}{c} \phi_1(p)\\ \phi_2(p)\end{array}\right) \,.
\end{equation}
We have the equations
\begin{eqnarray}\label{bispinor2}
(p_0+g_2p_0^2-(\vec{p}\cdot\vec{\sigma} ))   \phi_2=-m\phi_1   \,,\\
(p_0-g_2p_0^2+(\vec{p}\cdot\vec{\sigma} ) )  \phi_1=-m\phi_2\,.
\end{eqnarray}
We find for the negative-energy solutions associated 
to $ \Lambda_+^2=0$
\begin{eqnarray}
	v^{(1)}( p)&=&\left(\begin{array}{c}   \sqrt{p_0+g_2p_0^2+|\vec{p}| } \xi^{(-)}(-\vec p)\\ 
	 - \sqrt{p_0-g_2p_0^2-|\vec{p}| }  \xi^{(-)}(-\vec p) \end{array}\right)_{p_0= \omega_1}  \,,
\nonumber \\
	V^{(1)}( p)&=&\left(\begin{array}{c}  \sqrt{p_0+g_2p_0^2+|\vec{p}| }\xi^{(-)}(-\vec p)\\ \label{v_1}
	- \sqrt{p_0-g_2p_0^2-|\vec{p}| }\xi^{(-)}(-\vec p) \end{array}\right)_{p_0= W_1}  \,.\nonumber  \\
\end{eqnarray}
and to $ \Lambda_-^2=0$
\begin{eqnarray}
	v^{(2)}(p)&=&\left(\begin{array}{c} \sqrt{p_0+g_2p_0^2-|\vec{p} |}  \xi^{(+)}(\vec p)\\ 
	- \sqrt{p_0-g_2p_0^2+|\vec{p} |}  \xi^{(+)}(\vec p) \end{array}\right)_{p_0= \omega_2}  \,,
\nonumber \\
	V^{(2)}( p)&=&\left(\begin{array}{c}  \sqrt{p_0+g_2p_0^2-|\vec{p} |}  \xi^{(+)}(\vec p)\\ \label{v_2}
-	\sqrt{p_0-g_2p_0^2+|\vec{p} |} \xi^{(+)}(\vec p) \end{array}\right)_{p_0= W_2}  \,.\nonumber  \\
\end{eqnarray}
We can write some relations satisfied by the spinors, which do not 
apart too much from the usual expressions. They are
\begin{eqnarray}\label{Inner_uv}
u^{s\dag}({p}) u^{r}({p})&=&2\omega_s\delta^{rs}\,,   \nonumber \\
v^{s\dag}({p})v^{r}({p})&=&2\omega_s \delta^{rs}\,,
\end{eqnarray}
and 
\begin{eqnarray}\label{Inner_UV}
U^{s\dag}({p}) U^{r}({p})&=&2W_s\delta^{rs}\,, 
 \nonumber \\ V^{s\dag}({p})V^{r}({p})&=&2W_s \delta^{rs}\,,
\end{eqnarray}
and for the fields $\bar u=u^{\dag} \gamma_0$ we have
\begin{eqnarray}\label{Innerbar_uv}
\bar{u}^{s}({p}) u^{r}({p})&=&2m\delta^{rs}\,  
  \nonumber ,\\ \bar{v}^{s}({p})v^{r}({p})&=&-2m \delta^{rs}\,,
\end{eqnarray}
and 
\begin{eqnarray}\label{Innerbar_UV}
\bar{U}^{s}({p}) U^{r}({p})&=&2m\delta^{rs}\,,
  \nonumber \\\bar{V}^{s}({p})V^{(r)}({p})&=&-2m \delta^{rs}\,,
\end{eqnarray}
where the indices run over $r,s =1,2$.
The detailed derivation of the spinors, together with their complete
 inner and outer product relations 
 are given in the Appendix~\ref{App:A}.
%...................................................................................
\section{Quantization}\label{sectionIII}
%...................................................................................
In this section, we focus on the quantization of the Lorentz-violating fermion 
model. We derive the Hamiltonian and the four-dimensional 
representation of the Feynman propagator. In the last 
section, we study microcausality preservation.
%...................................................................................
\subsection{ETCR of the fields}\label{subsectionIII}
%...................................................................................
The Lagrangian~\eqref{HO_Lag} can be integrated by parts to produce
\begin{align}\label{sym_Lag}
\mathcal{L}'&=\frac{i}{2}(\psi^\dagger\dot{\psi}-\dot{\psi}^\dagger\psi)
+\bar{\psi}(i\gamma^i\partial_i-m)\psi
\notag \\ &- g_2\dot{\psi}^\dagger\gamma_5\dot{\psi} \,.  
\end{align}		  
The above Lagrangian~\eqref{sym_Lag} is equivalent to the original one, but it is 
simpler in the sense of being standard-derivative order and symmetrical with respect to
 time-derivatives. We work with this Lagrangian in the next sections.

It is convenient to decompose the field $\psi(\vec{x},x_0)$ 
in terms of two fields $\psi_1$ and $\psi_2$ as
\begin{eqnarray}\label{fields12}
\psi(\vec{x},x_0)=\psi_1(\vec{x},x_0)+\psi_2(\vec{x},x_0)  \,.
\end{eqnarray}
We take the field $\psi_1$ to describe 
standard particle states, which eventually includes perturbative 
corrections in the parameter $g_2$. 
On the other hand, the field $\psi_2$ is defined 
to be associated with negative-metric particles or ghosts.

We expand each field considering their
plane wave and spinor solutions found earlier. 
The particle field is
\begin{eqnarray}\label{Psi1}
\psi_1(\vec{x},x_0)&=&\sum_{r=1,2}\int\frac{d^3\vec p}{(2\pi)^3}  \frac{1}{\sqrt {{N_r}}}  \left(  a^{r}_p  u^{r}(p) 
 e^{-ip\cdot x}  \nonumber \right. \\ &+&\left.   b^{r\dagger}_p  v^{r}(p)     e^{ip\cdot x} \right)_{p_0=\omega_r}  \,,
\end{eqnarray}
and the ghost field
\begin{eqnarray}\label{Psi2}
\psi_2(\vec{x},x_0)&=&\sum_{r=1,2}\int\frac{d^3\vec p}{(2\pi)^3}  \frac{1}{\sqrt {   {\mathcal N_r} }} \left(  \alpha^{r}_p   U^{r}(p) 
 e^{-ip\cdot x}  \nonumber \right. \\ &+&\left.   \beta^{r  \dagger}_p V^{r}(p)    e^{ip\cdot x} \right)_{p_0=W_r}  \,.
\end{eqnarray}
We have introduced the creation operators $a^{\dag r}_p, b^{\dag r}_p$ 
and the annihilation operators
${a^r_p}, b^r_p$ for particle states and the set of operators
$ {\alpha^{\dag r}_p},{\beta^{\dag r}_p}$ and $ {\alpha^r_p},{\beta^r_p}$ representing 
creation and annihilation operators, respectively, for ghosts.

The fields $\psi_1(\vec{x},x_0)$ and $\psi_2(\vec{x},x_0)$ are normalized with the constants
\begin{eqnarray}\label{Norm_part}
{N_1}&=&2\omega_1   g_2^2 \left(    W^2_1-\omega^2_1\right)   \,,  \nonumber   \\ 
{N_2}&=&   2\omega_2   g_2^2 \left(   W^2_2-  \omega^2_2\right)   \,,  
\end{eqnarray}
and 
\begin{eqnarray}\label{Norm_neg}
{\mathcal N_1}&=& 2W_1   g_2^2\left(    W^2_1-\omega^2_1\right) \,,  \nonumber  \\
{\mathcal N_2}&=&  2W_2   g_2^2\left(    W^2_2- \omega^2_2\right)   \,.
\end{eqnarray}
In the Appendix~\ref{App:A}, we explain how they appear 
associated to a modified internal product between 
spinor states of positive and negative energy.

From the Lagrangian~\eqref{sym_Lag}, we compute the 
momenta associated to 
the independent fields $\psi$ and $\psi^\dagger$,
\begin{eqnarray}\label{mom_rel}
\pi_{\psi}&=&\frac{\partial \mathcal{L'}}{\partial \dot{\psi}}=\frac{i}{2}\psi^\dagger
-g_2\dot{\psi}^\dagger \gamma_5\,, \\ \label{mom_rel2}
 \pi_{\psi^{\dag}}&=&\frac{\partial \mathcal{L'}}{\partial {\dot \psi}^{\dag} 
  }=-\frac{i}{2}\psi-g_2\gamma_5 {\dot \psi} \,.
\end{eqnarray}
We impose the equal-time anticommutation relations for the fields and their conjugate momenta fields
\begin{eqnarray}\label{REP}
 \lbrace  {\psi}(\vec{x},x_0), {\pi_{\psi}} (\vec{y},x_0)\rbrace&=& i \delta^{(3)}(\vec{x}-\vec{y})\,, 
 \end{eqnarray}
\begin{eqnarray} \label{REP2}
  \lbrace  {\psi^{\dag}}(\vec{x},x_0), \pi_{\psi^{\dag} }(\vec{y},x_0)\rbrace&=& i \delta^{(3)}(\vec{x}-\vec{y})\,,
\end{eqnarray}
with the rest of commutators being zero.
In order to achieve Eqs.~\eqref{REP} and~\eqref{REP2}
we take the creation and annihilation operators to obey the rules
\begin{eqnarray}\label{Alg_pos}
\lbrace {a}_p^{s},{a}_k^{r\dagger}\rbrace&=&(2\pi)^3\delta^{sr}
\delta^{(3)}(\vec{k}-\vec{p})\,,  \notag \\  \lbrace {b}_p^{s},
{b}_k^{r\dagger}\rbrace&=&(2\pi)^3\delta^{sr}\delta^{(3)}(\vec{k}-\vec{p}) \,,
\end{eqnarray}
and 
\begin{eqnarray}\label{Alg_neg}
\lbrace {\alpha}_p^{s},{\alpha}_k^{r\dagger}\rbrace&=&-(2\pi)^3\delta^{sr}
\delta^{(3)}(\vec{k}-\vec{p})\,, \notag \\    \label{Alg_neg2} \lbrace {\beta}_p^{s},{\beta}_k^{r\dagger}
\rbrace&=&-(2\pi)^3\delta^{sr}\delta^{(3)}(\vec{k}-\vec{p})\,,
\end{eqnarray}
with the vacuum defined by 
\begin{eqnarray}
{a}_p^{s}    \ket{0} =  {b}_p^{s}  \ket{0}  ={\alpha}_p^{s}  \ket{0}=  {\beta}_p^{s}  \ket{0}=0\,.
\end{eqnarray}
Notice that the second set of rules are defined with a nonstandard 
negative sign in~\eqref{Alg_neg} 
which is the first indication of having an indefinite metric in Hilbert space.

In fact, we can write down the metric for each sector
in the indefinite Hilbert space.
We define the $n-$particle states of polarization $s$ to appear
by applying repeatedly creation operators on the vacuum state. For particles states
\begin{align}
\ket{n_{1,s}}=\frac{1}{\sqrt{(n_{1,s})!}}  ({a}_p^{s\dag } )^{n_{1,s}}\ket{0} \,,
\end{align}
and for ghost states
\begin{align}
\ket{n_{2,s}}=\frac{1}{\sqrt{(n_{2,s)}!}}  ({\alpha}_p^{s\dag } )^{n_{2,s}}\ket{0}\,,
\end{align}
where $n_{1,s}$ and $n_{2,s}$ are the eigenvalues of the number operators 
$\hat N_{1,s}=  {a}_p^{s\dag }   {a}_p^{s }$
and $\hat N_{2,s}=  {\alpha}_p^{s\dag }   {\alpha}_p^{s }$, respectively.
Hence, for particles we have the positive-metric
\begin{align}
\eta_{1,s}=   \bra{n_{1,s}}   \ket{n_{1,s}}=1\,,
\end{align}
and for ghost states the indefinite-metric 
\begin{align}
 \eta_{2,s}= \bra{n_{2,s}}   \ket{n_{2,s}} =(-1)^{n_{2,s}}\,.
\end{align}
From~\eqref{Psi1} and~\eqref{Psi2} we have
\begin{align}
\psi^{\dag} (\vec{x},x_0)=\psi_1^{\dag} (\vec{x},x_0)+\psi_2^{\dag}(\vec{x},x_0) \,,
\end{align}
where 
\begin{align}
\psi_1^{\dag}(\vec{x},x_0)&=\sum_{r=1,2}\int\frac{d^3\vec p}{(2\pi)^3} 
 \frac{1}{\sqrt {{N_r}}}  \left(  a^{r\dag}_p  u^{r\dag}(p) 
 e^{ip\cdot x}  \nonumber \right. \\ &+\left.   b^{r}_p  v^{r\dag}(p)   
   e^{-ip\cdot x} \right)_{p_0=\omega_r}  \,,
\\
\psi_2^{\dag}(\vec{x},x_0)&=\sum_{r=1,2}\int\frac{d^3\vec p}{(2\pi)^3} 
\frac{1}{\sqrt {   {\mathcal N_r} }} \left(  \alpha^{r\dag }_p   U^{r\dag}(p) 
 e^{ip\cdot x}  \nonumber \right. \\ &+\left.   \beta^{r}_p V^{r\dag}(p)    e^{-ip\cdot x} \right)_{p_0=W_r}  \,.
\end{align}
We introduce momenta with respect to the decomposed fields in the form
\begin{eqnarray} \label{newcomm1}
\pi_{1}&=&\frac{\partial \mathcal{L'}}{\partial \dot{\psi}_1}=\frac{i}{2}
\psi_1^\dagger-g_2\dot{\psi}_1^\dagger \gamma_5 \,,
\\ \label{newcomm2}
\pi_{2}&=&\frac{\partial \mathcal{L'}}{\partial \dot{\psi}_2}=\frac{i}{2}
\psi_2^\dagger-g_2\dot{\psi}_2^\dagger \gamma_5 \,,
\end{eqnarray}
and
\begin{eqnarray}		
\pi_1^{\dagger}&=&\frac{\partial \mathcal{L'}}{\partial \dot{\psi_1}^\dagger}
=-\frac{i}{2}\psi_1-g_2\gamma_5\dot{\psi_1}\,,
\\		
\pi_2^{\dagger}&=&\frac{\partial \mathcal{L'}}{\partial \dot{\psi_2}^\dagger}
=-\frac{i}{2}\psi_2-g_2\gamma_5\dot{\psi_2}\,.
\end{eqnarray}
Therefore, we can write
\begin{align}
\pi_{\psi}&=\pi_{1}+\pi_{2}\,, 
\\
 \pi_{\psi^{\dag}}&=\pi_1^{\dagger}+\pi_2^{\dagger} \,.
\end{align}
With these simplifications, we start computing the commutator~\eqref{REP}. 
We can write the first commutator as the sum 
\begin{align}\label{ETCrel}
    \lbrace \psi(\vec x,x_0),\pi(\vec y,x_0)\rbrace& =\lbrace  \psi_1(\vec x,x_0),\pi_1(\vec y,x_0)\rbrace 
\notag \\ &+ \lbrace \psi_2(\vec x,x_0),\pi_2(\vec y,x_0)\rbrace   \,, 
\end{align}
and momenta~\eqref{newcomm1} and \eqref{newcomm2} as
\begin{align} 
{\pi}_1(\vec{x},x_0)&=i\sum_{s}\int\frac{d^3\vec p}{(2\pi)^3}   \frac{1}{\sqrt N_s} \left[ {a}_p^{s\dagger} 
u^{s\dagger}(p)  \left(\frac{1}{2}-g_2\omega_s\gamma_5\right)     
  \right. \nonumber \\ 
&\times    \left.    e^{ip\cdot x}  + {b}^{s}_p  v^{s\dagger}(p)
 \left(\frac{1}{2}+g_2\omega_{s}\gamma_5\right)   e^{-ip\cdot x}
   \right] _{p_0=\omega_s}\,,
\end{align}
and
\begin{align} 
{\pi}_2(\vec{x},x_0)&=i\sum_{s}\int\frac{d^3\vec p}{(2\pi)^3}   \frac{1}{\sqrt
 {\mathcal N}_s} \left[{\alpha}_p^{s\dagger} U^{s\dagger}(p) 
\left(\frac{1}{2}-g_2W_s\gamma_5\right)        \right. \nonumber \\ 
&\times    e^{ip\cdot x} \left.+ {\beta}^{s}_p  V^{s\dagger}(p) \left(\frac{1}{2}
+g_2W_{s}\gamma_5\right)   e^{-ip\cdot x}
   \right]_{p_0=W_s}\,. 
\end{align}
The first commutator in \eqref{ETCrel} can be shown to be
\begin{align}
  &  \lbrace \psi_1(\vec x,x_0),\pi_1(\vec y,x_0)\rbrace \nonumber  \\ &=\sum_{r=1,2}\int 
  \frac{d^3\vec p}{(2\pi)^3}\frac{i}{N_r}\left[ u^r(p)u^{r\dagger}(p)\left(\frac{1}{2}-g_2\omega_r\gamma_5\right) \right. \nonumber  \\
  &+  \left. v^r(-p)v^{r\dagger}(-p)\left(\frac{1}{2}+g_2\omega_r\gamma_5\right)\right]   e^{i\vec p\cdot(\vec x-\vec y)} \,,
\end{align}
We can proceed analogously and by considering the minus sign due to the minus in 
the anticommutation relations~\eqref{Alg_neg} we obtain
\begin{align}
  &  \lbrace \psi_2(\vec x,x_0),\pi_2(\vec y,x_0)\rbrace \nonumber  \\ &=-\sum_{r=1,2}\int \frac{d^3\vec p}{(2\pi)^3}\frac{i}{N_r}
  \left[ U^r(p)U^{r\dagger}(p)\left(\frac{1}{2}-g_2\omega_r\gamma_5\right) \right. \nonumber  \\
  &+  \left. V^r(-p)V^{r\dagger}(-p)\left(\frac{1}{2}+g_2\omega_r\gamma_5\right)\right]   e^{i\vec p\cdot(\vec x-\vec y)} \,.
\end{align}
We use Eqs.~\eqref{uu1+vv1dag}~\eqref{uu1-vv1dag},~\eqref{uu2+vv2dag} and~\eqref{uu2-vv2dag}
given in the Appendix~\eqref{subappendixC}.
We arrive at
\begin{align}\label{C1}
  & \lbrace \psi_1(\vec x,x_0),\pi_1(\vec y,x_0)\rbrace=i\int \frac{d^3\vec p}{(2\pi)^3}\left(\frac{\omega_1}{N_1}
\left[\frac{1}{2}     (\mathbb{1}_{4}-Q)   \right.  \right.     \notag   \\   & \left.  \left. 
-g_2(\gamma^ip_i+m-g_2\omega_1^2\gamma_0\gamma_5)\gamma_0   (\mathbb{1}_{4}-Q)   \gamma_5\right]  \right. 
\nonumber   \\
 &\left. +\frac{\omega_2}{N_2}\left[ \frac{1}{2}   (\mathbb{1}_{4}+Q)      
    -g_2(\gamma^ip_i+m              \right.  \right. \notag \\   & \left.  \left.        
        -g_2\omega_2^2\gamma_0\gamma_5)\gamma_0
 (\mathbb{1}_{4}+Q)    \gamma_5\right]\right) 
e^{i\vec p\cdot(\vec x-\vec y)} \,,
\end{align}
and to
\begin{align}\label{C2}
& \lbrace \psi_2(\vec x,x_0),\pi_2(\vec y,x_0)\rbrace=-i\int \frac{d^3\vec p}
{(2\pi)^3}\left(\frac{W_1}{\mathcal{N}_1}\left[\frac{1}{2}    (\mathbb{1}_{4}-Q)    \right.  \right.  \notag  \\   
&  \left.    \left. -g_2(\gamma^ip_i+m-g_2W_1^2
\gamma_0\gamma_5)\gamma_0   (\mathbb{1}_{4}-Q)   \gamma_5\right]     \right.      \notag   \\
& \left.    +\frac{W_2}{\mathcal{N}_2}\left[ \frac{1}{2}     (\mathbb{1}_{4}+Q)     
   -g_2(\gamma^ip_i+m       \right.  \right. \notag \\   & \left.  \left.          
    -g_2W_2^2\gamma_0\gamma_5)\gamma_0
 (\mathbb{1}_{4}+Q) 
\gamma_5\right]\right)     e^{i\vec p\cdot(\vec x-\vec y)} \,.
\end{align}
We use the relations
\begin{align}
    \frac{\omega_1}{N_1}=\frac{W_1}{\mathcal N_1}= \frac{1}{2g_2^2(W_1^2-\omega_1^2)}\,,
\end{align}
and by adding~\eqref{C1} and~\eqref{C2} produces
\begin{align}
  &  \lbrace \psi(\vec x,x_0),\pi(\vec y,x_0)\rbrace
   =i\int\frac{d^3\vec p}{(2\pi)^3}  \left[ \frac{\gamma_0\gamma_5\gamma_0}{2g_2^2(W_1^2-\omega_1^2)} 
        \right.  \nonumber   \\ &\times \left.  \left(  g_2^2 (\omega_1^2-W_1^2) 
  (\mathbb{1}_{4}-Q)   \gamma_5
   \right)   \right. \nonumber \\
    &+ \left.  \frac{\gamma_0\gamma_5\gamma_0}{2g_2^2(W_2^2-\omega_2^2)}\left(g_2^2(\omega_2^2-W_2^2)
     \nonumber  \right. \right.  \\ &\times \left.  \left.     (\mathbb{1}_{4}+Q)    
        \gamma_5\right) \right]  e^{i\vec p\cdot(\vec x-\vec y)}\,, 
 \end{align}   
or
\begin{align}
& \lbrace \psi(\vec x,x_0),\pi(\vec y,x_0)\rbrace =-i\int
\frac{d^3\vec p}{(2\pi)^3}\left(\frac{1}{2}\gamma_0\gamma_5
\gamma_0   \nonumber  \right.  \\     \times &   (\mathbb{1}_{4}-Q)     \gamma_5+   \left. \frac{1}{2}
\gamma_0\gamma_5\gamma_0    (\mathbb{1}_{4}+Q)   \gamma_5\right)e^{i\vec p\cdot(\vec x-\vec y)} \,. 
\end{align}   
Finally
\begin{eqnarray}
   \lbrace \psi(\vec x,x_0),\pi(\vec y,x_0)\rbrace 
   &=&-i\int\frac{d^3\vec p}{(2\pi)^3}(\gamma_0\gamma_5\gamma_0\gamma_5)
   e^{i\vec p\cdot(\vec x-\vec y)} \nonumber \\
    &=&i\delta^{(3)}(\vec x-\vec y)\,.
\end{eqnarray}
In a similar way the commutator~\eqref{REP2} is also satisfied.
%...................................................................................
\subsection{The Hamiltonian}
%...................................................................................
The Legendre transformation of the Lagrangian~\eqref{sym_Lag}
produces the Hamiltonian
\begin{eqnarray}
H&=& \int d^3\vec x  \left( \pi_\psi \dot{\psi}+\dot{\psi}^\dagger\pi_{{\psi}^\dagger}-\mathcal{L}' \right)\,.
\end{eqnarray}
Considering momenta in~Eqs.~\eqref{mom_rel} and~\eqref{mom_rel2} the Hamiltonian can be cast into the form
\begin{equation}
H=\int d^3\vec x\left( -g_2\dot{\psi}^\dagger \gamma_5\dot{\psi} +\bar{\psi}(-i\gamma^i\partial_i+m)\psi    \right)\,.
\end{equation}
With the decomposition of fields~\eqref{fields12} let us write
\begin{eqnarray}
H  &\equiv & \sum _{a,b=1,2} H_{ab}=\sum _{a,b=1,2}\int d^3\vec x \;  \mathcal{H}_{ab}(x)  \,,
\end{eqnarray}
where 
\begin{align}\label{Ham_inparts}
 \mathcal{H} _{ab} (x)&=  -g_2\dot{\psi}_a^\dagger   (x)\gamma_5\dot{\psi}_b(x)    \notag  \\   
&
 +\bar{\psi}_a(x)  (-i\gamma^k\partial_k+m)\psi_b(x)  \,.
\end{align} 
We write the contributions coming from both fields separately. 

The contributions coming from $\psi_1$ are
\begin{align}
&-g_2\gamma_5\dot{\psi}_1=  -g_2\gamma_5  \sum_{s}\int\frac{d^3p'}{(2\pi)^3   }     \frac{1}{\sqrt {{N'_s}}}    \left( 
(-i\omega'_s)        \right. \notag  \\   
&\times  \left.    u^s(p')  {a}_{p'}^s     e^{-ip'\cdot x}  
+    (i\omega'_{s})  v^s(p')  {b}^{s\dagger}_{p'}   e^{ip'\cdot x}  
\right)_{ p'_0=\omega'_s }
\end{align}
and
\begin{align} \label{second_Term1}
&(-i\gamma^i\partial_i+m)\psi _1(x) =\sum_{s}\int\frac{d^3p'}{(2\pi)^3}  \frac{1}{\sqrt {{N'_s}}}     \left(    
(-\gamma^i p'_i+m)      \right. \notag  \\   
&\times  \left.    u^s(p')  {a}_{p'}^s  e^{-ip' \cdot x}     +  (\gamma^i  p'_i+m)v^s(p')   {b}^{s\dagger}_{p'}  
  e^{ip'\cdot x}      \right)    _{p'_0=\omega'_s}\,,
\end{align}
And the ones coming from $\psi_2$ are
\begin{align}
&-g_2\gamma_5\dot{\psi}_2=  -g_2\gamma_5  \sum_{s}\int\frac{d^3p'}{(2\pi)^3   }     \frac{1}{\sqrt {{\mathcal{N}'_s}}}    \left( 
(-iW'_s)        \right. \notag  \\   
&\times  \left.    U^s(p')  {\alpha}_{p'}^s     e^{-ip'\cdot x}  
+    (iW'_{s})  V^s(p')  {\beta}^{s\dagger}_{p'}   e^{ip'\cdot x}  
\right)_{ p'_0=W'_s }\,,
\end{align}
and
\begin{align}\label{second_Term2}
&(-i\gamma^i\partial_i+m)\psi _2(x) =\sum_{s}\int\frac{d^3p'}{(2\pi)^3}  \frac{1}{\sqrt {{\mathcal{N}'_s}}}     \left(    
(-\gamma^i p'_i+m)      \right. \notag  \\   
&\times  \left.    U^s(p')  {\alpha}_{p'}^s  e^{-ip' \cdot x}     +  (\gamma^i  p'_i+m)V^s(p')   {\beta}^{s\dagger}_{p'}  
  e^{ip'\cdot x}      \right)    _{p'_0=W'_s}\,,
\end{align}
We can rewrite the second terms~\eqref{second_Term1} and~\eqref{second_Term2}
using the 
equations of motion~\eqref{eqM} and~\eqref{neg_sol}, i.e., 
\begin{align}
(-\gamma^i p'_i+m) u^s(p')&=\gamma_0 ( \omega'_s-g_2 \gamma_5 \omega_s^{\prime2})u^s(p') \,,
\notag \\
(\gamma^i p'_i+m) v^s(p')&= - \gamma_0  ( \omega'_s+g_2 \gamma_5 \omega_s^{\prime 2})v^s(p') \,.
\end{align} 
and
\begin{align}
(-\gamma^i p'_i+m) U^s(p')&=\gamma_0 ( W'_s-g_2 \gamma_5 W_s^{\prime 2})U^s(p')
\,,   \notag \\
(\gamma^i p'_i+m) V^s(p')&=- \gamma_0 ( W'_s+g_2 \gamma_5 W_s^{\prime 2})V^s(p')
\,.
\end{align} 
This yields 
\begin{align}
&(-i\gamma^i\partial_i+m)\psi_1(x)=\sum_{s}\int\frac{d^3\vec p'}{(2\pi)^3}   
  \frac{1}{\sqrt {{{N}'_s}}}    \notag \\ &\times    \left[\left(
 \gamma_0(\omega'_s-g_2\omega_s^{\prime 2}\gamma_5)u^s(p') {a}_{p'}^s
 e^{-i\omega'_sx_0}   \right)   e^{i\vec{p'}\cdot\vec{x}}\right.  \notag \\
		&-\left.\left(    \gamma_0(\omega'_{s}+g_2\omega_{s}^{\prime 2}
		\gamma_5)v^s(p') {b}^{s\dagger}_{p'}e^{i\omega'_{s}x_0}  \right)    e^{-i\vec{p'}\cdot\vec{x}} \right]
\end{align}
and
\begin{align}
&(-i\gamma^i\partial_i+m)\psi_2(x)= \sum_{s}\int\frac{d^3\vec p'}{(2\pi)^3}  
 \frac{1}{\sqrt {{\mathcal{N}'_s}}}  \notag \\ &\times   \left[\left( \gamma_0(W'_s-g_2W_s^{\prime 2}
 \gamma_5)U^s(p'){\alpha}^s_{p'}e^{-iW'_sx_0}\right)e^{i\vec{p'}\cdot\vec{x}}\right.  \notag \\
&-\left.\left(     \gamma_0(W'_{s}+g_2W_{s}^{\prime 2}\gamma_5)V^s(p')
{\beta}^{s\dagger}_{p'}e^{iW'_{s}x_0   }\right)    e^{-i\vec{p'}\cdot\vec{x}} \right]
\end{align}
Now, it is convenient to decompose further by considering
\begin{align}
H_{11}&=H^{uu}+H^{uv}+H^{vv}+H^{vu}   \,,
\notag   \\
H_{12}&=H^{uU}+H^{uV}+H^{vU}+H^{vV}   \,,
\notag \\
H_{21}&=H^{Uu}+H^{Uv}+H^{Vu}+H^{Vv}   \,,
\notag \\
H_{22}&=H^{UU}+H^{UV}+H^{VU}+H^{VV}   \,,
\end{align}	
After some algebra we find the particle contributions
\begin{align}
{H}^{uu}&= \sum_{r,s} \int\frac{d^3\vec p}{(2\pi)^3}  \frac{1}{   { \sqrt{N_rN_s}} }  
a^{r\dag }_p a^{s }_p  \, e^{i(\omega_r -\omega_s) x_0}
 \notag \\ &\times \omega_s  u^{r \dag }(p)  ( 1-g_2\gamma_5( \omega_s+ \omega_r )  ) 
  u^{s}(p)  \,,
\\
{H}^{uv}&= -\sum_{r,s} \int\frac{d^3\vec p}{(2\pi)^3}  \frac{1}{   { \sqrt{N_rN_s}} } 
 a^{r\dag }_p b^{s\dag }_{-p} \, e^{i(\omega_r+\omega_s) x_0}
 \nonumber \\ &\times   \omega_s   u^{r \dag }(p) ( 1+g_2\gamma_5( \omega_s- \omega_r )  )  
    v^{s}(-p)  \,,
\\
{H}^{vu}&= \sum_{r,s} \int\frac{d^3\vec p}{(2\pi)^3} \frac{1}{   { \sqrt{N_rN_s}} } 
 b^{r }_p a^{s }_{-p}  \, e^{-i(\omega_r+\omega_s) x_0}
 \nonumber \\ &\times \omega_s   v^{r \dag }(p)( 1-g_2\gamma_5( \omega_s- \omega_r )  ) 
   u^{s}(-p)  \,,
\\
{H}^{vv} &= -\sum_{r,s} \int\frac{d^3\vec p}{(2\pi)^3} \frac{1}{   { \sqrt{N_rN_s}} } 
b^{r}_p b^{s\dag }_p \, e^{-i(\omega_r-\omega_s) x_0}
 \nonumber \\ &\times    \omega_s   v^{r \dag }(p)  ( 1+g_2\gamma_5
 ( \omega_s+ \omega_r )  ) v^{s}(p)  \,,
\end{align} 
the mixed ones
\begin{align}
{H}^{uU}&= \sum_{r,s} \int\frac{d^3\vec p}{(2\pi)^3}  \frac{1}{   { \sqrt{N_r  \mathcal{N}_s}} }  
a^{r\dag }_p \alpha^{s }_p  \, e^{i(\omega_r -W_s) x_0}
 \notag \\ &\times W_s  u^{r \dag }(p)  ( 1-g_2\gamma_5( W_s+ \omega_r )  ) 
  U^{s}(p)  \,,
\\
{H}^{uV} &= -\sum_{r,s} \int\frac{d^3\vec p}{(2\pi)^3}  \frac{1}{   { \sqrt{N_r  \mathcal{N}_s}} } 
 a^{r\dag }_p \beta^{s\dag }_{-p} \, e^{i(\omega_r+W_s) x_0}
 \nonumber \\ &\times  W_s   u^{r \dag }(p) ( 1+g_2\gamma_5( W_s- \omega_r )  )  
    V^{s}(-p)  \,,
\\
{H}^{vU}&= \sum_{r,s} \int\frac{d^3\vec p}{(2\pi)^3} \frac{1}{   { \sqrt{N_r   \mathcal{N}_s}} } 
 b^{r }_p \alpha^{s }_{-p}  \, e^{-i(\omega_r+W_s) x_0}
 \nonumber \\ &\times W_s   v^{r \dag }(p)( 1-g_2\gamma_5( W_s- \omega_r )  ) 
   U^{s}(-p)  \,,
\\
{H}^{vV}&= -\sum_{r,s} \int\frac{d^3\vec p}{(2\pi)^3} \frac{1}{   { \sqrt{N_r  \mathcal{N}_s}} } 
b^{r}_p \beta^{s\dag }_p \, e^{-i(\omega_r-W_s) x_0}
 \nonumber \\ &\times   W_s   v^{r \dag }(p)  ( 1+g_2\gamma_5
 ( W_s+ \omega_r )  ) V^{s}(p)  \,,
\end{align} 
\begin{align}
{H}^{Uu}&= \sum_{r,s} \int\frac{d^3\vec p}{(2\pi)^3}  \frac{1}{   { \sqrt{\mathcal{N}_rN_s}} }  
\alpha^{r\dag }_p a^{s }_p  \, e^{i(W_r -\omega_s) x_0}
 \notag \\ &\times \omega_s  U^{r \dag }(p)  ( 1-g_2\gamma_5( \omega_s+ W_r )  ) 
  u^{s}(p)  \,,
\\
{H}^{Uv}&= -\sum_{r,s} \int\frac{d^3\vec p}{(2\pi)^3}  \frac{1}{   { \sqrt{ \mathcal{N}_rN_s}} } 
 \alpha^{r\dag }_p b^{s\dag }_{-p} \, e^{i(W_r+\omega_s) x_0}
 \nonumber \\ &\times   \omega_s   U^{r \dag }(p) ( 1+g_2\gamma_5( \omega_s-W_r )  )  
    v^{s}(-p)  \,,
\\
{H}^{Vu} &= \sum_{r,s} \int\frac{d^3\vec p}{(2\pi)^3} \frac{1}{   { \sqrt{\mathcal{N}_rN_s}} } 
 \beta^{r }_p a^{s }_{-p}  \, e^{-i(W_r+\omega_s) x_0}
 \nonumber \\ &\times \omega_s   V^{r \dag }(p)( 1-g_2\gamma_5( \omega_s- W_r )  ) 
   u^{s}(-p)  \,,
\\
{H}^{Vv}&= -\sum_{r,s} \int\frac{d^3\vec p}{(2\pi)^3} \frac{1}{   { \sqrt{ \mathcal{N}_rN_s}} } 
\beta^{r}_p b^{s\dag }_p \, e^{-i(W_r-\omega_s) x_0}
 \nonumber \\ &\times    \omega_s   V^{r \dag }(p)  ( 1+g_2\gamma_5
 ( \omega_s+W_r )  ) v^{s}(p)  \,,
\end{align} 
and the ghost contributions
\begin{align}
{H}^{UU}&= \sum_{r,s} \int\frac{d^3\vec p}{(2\pi)^3}  \frac{1}{   { \sqrt{\mathcal{N}_r\mathcal{N}_s}} }  
\alpha^{r\dag }_p \alpha^{s }_p  \, e^{i(W_r -W_s) x_0}
 \notag \\ &\times W_s  U^{r \dag }(p)  ( 1-g_2\gamma_5( W_s+ W_r )  ) 
  U^{s}(p)  \,,
\\
{H}^{UV}&= -\sum_{r,s} \int\frac{d^3\vec p}{(2\pi)^3}  \frac{1}{   { \sqrt{\mathcal{N}_r  \mathcal{N}_s}} } 
 \alpha^{r\dag }_p \beta^{s\dag }_{-p} \, e^{i(W_r+W_s) x_0}
 \nonumber \\ &\times W_s   U^{r \dag }(p) ( 1+g_2\gamma_5( W_s- W_r )  )  
    V^{s}(-p)  \,,
\\
{H}^{VU}&= \sum_{r,s} \int\frac{d^3\vec p}{(2\pi)^3} \frac{1}{   { \sqrt{\mathcal{N}_r  \mathcal{N}_s}} } 
 \beta^{r }_p \alpha^{s }_{-p}  \, e^{-i(W_r+W_s) x_0}
 \nonumber \\ &\times W_s   V^{r \dag }(p)( 1-g_2\gamma_5( W_s- W_r )  ) 
   U^{s}(-p)  \,,
\\
{H}^{VV} &= -\sum_{r,s} \int\frac{d^3\vec p}{(2\pi)^3} \frac{1}{   { \sqrt{  \mathcal{N}_r \mathcal{N}_s}} } 
\beta^{r}_p \beta^{s\dag }_p \, e^{-i(W_r-W_s) x_0}
 \nonumber \\ &\times    W_s   V^{r \dag }(p)  ( 1+g_2\gamma_5
 ( W_s+ W_r )  ) V^{s}(p)  \,.
\end{align} 
After considering the sixteen terms and 
using the equations~\eqref{Prod_uu}~\eqref{Prod_UU} and~\eqref{Mixed} 
of the Appendix~\eqref{App:A}
the only non-zero contributions are
\begin{align}
{H}^{uu} &= \sum_{s}  \int\frac{d^3\vec p}{(2\pi)^3}  \omega_s a^{s\dag }_p a^{s }_p   \,,
\notag \\
{H}^{vv} &=- \sum_{s}  \int\frac{d^3\vec p}{(2\pi)^3}  \omega_s b^{s }_p b^{s \dag}_p   \,.
\end{align} 
and 
\begin{align}
{H}^{UU} &= -\sum_{s}  \int\frac{d^3\vec p}{(2\pi)^3}  W_s \alpha^{s\dag }_p \alpha^{s }_p  \,,
\notag \\
{H}^{VV} &=\sum_{s}  \int\frac{d^3\vec p}{(2\pi)^3}  W_s \beta^{s}_p \beta^{s \dag}_p   \,.
\end{align} 
Finally, adding all the parts we arrive at
\begin{align}
H&=\sum_{s=1,2}\int \frac{d^3\vec p}{(2\pi)^3}  \left( \omega_s {a}^{s\dagger}_p {a}_p^s
 -  \omega_{s} {b}^{s}_{p}{b}^{s\dagger}_{p}  
        \right.  \nonumber \\  &- \left.  W_s
{\alpha}^{s\dagger}_p{\alpha}_p^s   +W_{s}
{\beta}_{p}^{s}\hat{\beta}^{s\dagger}_{p} \right)\,,
\end{align}
and the normal ordering gives 
\begin{eqnarray}
:{H}:&=&\sum_{s=1,2}\int \frac{d^3p}{(2\pi)^3}\left(\omega_s ({a}^{s\dagger}_p \hat{a}_p^s
 +{b}^{s\dagger}_{p}{b}^{s}_{p}  
  )\right.  \nonumber 
\\&& \left.-W_s(
{\alpha}^{s\dagger}_p  {\alpha}_p^s+
{\beta}^{s\dagger}_{p}{\beta}_{p}^{s}  ) \right)\,.
\end{eqnarray}
The Hamiltonian is stable and in the presence of interaction we can always 
redefine the vacuum in order to produce 
a well bounded Hamiltonian. For fermions this is always possible due to the
 invariance of 
 the algebra~\eqref{Alg_neg} under a vacuum redefinition~\cite{LW}. However, 
 it is noted that for energies higher than $\frac{1}{2g_2}\sqrt{1+4m^2g_2^2}$
 at which the solutions $\pm \omega_1$ and $\pm W_1$ become complex, the 
 Hamiltonian is no longer hermitian.
%..................................................................................................
\subsection{ The Feynman propagator   }\label{subsectionV}
%.................................................................................................
We compute the modified
propagator starting from its definition 
\begin{align}
S_F&(x-y)=\bra{0}T\lbrace \psi(x),\bar\psi(y)\rbrace \ket{0} \,,
\end{align}
and in terms of theta functions and vacuum expectation values of fields we have
\begin{align}
S_F(x-y)&=\theta(x_0-y_0)\bra{0} \psi(x)\bar\psi(y)\ket{0}  \notag \\ &
-\theta(y_0-x_0) \bra{0} \bar\psi(y)\psi(x)\ket{0} \,.
\end{align}
To simplify the calculation and without loss of generality we set
$y = 0$.

We start with 
the case $x_0 > 0$ and define
\begin{align}
S_F(x)&=S^{(>)}_F(x)\equiv \bra{0} \psi(x)\bar\psi(0)\ket{0} \,.
\end{align}
Using the decomposition of fields in~Eq.\eqref{fields12} we can write
\begin{align}
S^{(>)}_F(x)=\bra{0} &\psi_1(x)\bar{\psi}_1(0) \ket{0}\notag
 \\ &+\bra{0} \psi_2(x)\bar{\psi}_2(0) \ket{0} \,.
\end{align}
Consider
\begin{align}
& \bra{0} \psi_1(x)\bar{\psi}_1(0) \ket{0}=\sum_{r,s=1,2}
\int\frac{d^3\vec p}{(2\pi)^3}\frac{d^3\vec k}{(2\pi)^3}
\notag \\ &  \bra{0}   \frac{1}{\sqrt {{N_r}}}  \left(  a^{r}_p  u^{r}(p)
 e^{-ip  x} + b^{r\dagger}_p  v^{r}(p)     e^{ip\cdot x} \right)_{p_0=\omega_r}   \notag \\
 &\times    \left( \frac{1}{\sqrt {{N_s}}}  \left(  a^{s\dagger}_k  \bar{u}^{s}(k) 
 + b^{s}_k  \bar{v}^{s}(k)      \right)_{k_0=\omega_s} \right) \ket{0} \,.
\end{align}
The action of the annihilation operators on the vacuum produces
\begin{align}
 & \bra{0} \psi_1(x)\bar{\psi}_1(0) \ket{0}=\sum_{r,s=1,2}\int\frac{d^3\vec p}
 {(2\pi)^3}\frac{d^3\vec k}{(2\pi)^3}  \frac{1}{\sqrt {{N_r}}\sqrt{N_s}}
\notag \\ &\times  u^{r}(p)  \bar{u}^{s}(k) \, \bra{0}a^{r}_p a^{s\dagger}_k\ket{0} 
 e^{-ip_r x}  \,,
\end{align}
 where $p_r=(\omega_r,\vec{p})$
and from the anticommutation relations~\eqref{Alg_pos} one has
\begin{align}
\bra{0} \psi_1(x)\bar{\psi}_1(0) \ket{0}&=\sum_{r=1,2}\int\frac{d^3
\vec p}{(2\pi)^3} \frac{1}{N_r}    u^{r}(p)  \bar{u}^{r}(p)
 e^{-ip_r x} \,.
 \end{align}
Now we use the expression~\eqref{uexp11} and~\eqref{uexp2} to arrive at
 \begin{align}
 & \bra{0} \psi_1(x)\bar{\psi}_1(0) \ket{0}\\ &=\int\frac{d^3\vec p}{(2\pi)^3}\left( 
    (\gamma_0\omega_1+\gamma^ip_i
 +m-g_2\omega_1^2\gamma_0\gamma_5)           \right.\nonumber \\
 &\left.    \frac 12 (\mathbb{1}_{4} -Q)  \,
 \frac{e^{-i\omega_1 x_0}}{N_1}   +   (\gamma_0\omega_2+\gamma^ip_i+m-
 g_2\omega_2^2\gamma_0\gamma_5) \right.  \notag 
  \\ &  \left.   \frac 12 (\mathbb{1}_{4} +Q)
 \frac{e^{-i\omega_2 x_0}    }{N_2}\right) e^{i\vec{p}\cdot\vec{x}} \,,
 \end{align}
we factorize the global operator
\begin{align}
 & \bra{0} \psi_1(x)\bar{\psi}_1(0) \ket{0}   \notag \\ &=(i\sla{\partial}+m+g_2
 \gamma_0\gamma_5\partial_0^2)
 \int\frac{d^3\vec p}{(2\pi)^3}\left[\frac{1}{2}  \left(\mathbb{1}_{4} -Q \right) \right.\nonumber \\
 &\times \left.   \frac{ e^{-i\omega_1 x_0}}{N_1} +\frac{1}{2}  \left(\mathbb{1}_{4} +Q \right)  \frac{
 e^{-i\omega_2 x_0}}{N_2} \right]e^{i\vec p\cdot\vec x}\,.
\end{align}
Analogously, for the ghost field we find
\begin{align}
\bra{0} \psi_2(x)\bar{\psi}_2(0) \ket{0}&=-(i\sla{\partial}
+m+g_2\gamma_0\gamma_5\partial_0^2)\notag \\ &\times \int\frac{d^3\vec p}
{(2\pi)^3}\left[    \frac{1}{2}  \left(\mathbb{1}_{4} -Q \right)   \frac{ e^{-iW_1 x_0}}{\mathcal{N}_1}
\right.\nonumber \\
 & \left.+ \frac{1}{2}  \left(\mathbb{1}_{4} +Q \right)   \frac{
 e^{-iW_2 x_0}}{\mathcal{N}_2} \right]e^{i\vec p\cdot\vec x}\,.
\end{align}
where a minus sign has appeared due to the ghost 
oscillators anticommutation relations.

Adding both contribution produces
\begin{align}\label{S>}
S_F^{(>)}(x)&=(i\slashed{\partial}+m+g_2\gamma_0\gamma_5
\partial_0^2)\int\frac{d^3\vec p}{(2\pi)^3}\left[   \frac{1}{2}  \left(\mathbb{1}_{4} -Q \right)
 \right.\nonumber \\
 &\times \left.    \left[\frac{ e^{-i\omega_1 x_0}}{N_1}-
 \frac{ e^{-iW_1 x_0}}{\mathcal{N}_1}\right]  + \frac{1}{2}  \left(\mathbb{1}_{4} +Q \right)
  \right. \notag  \\ &  \left. \left[ \frac{
 e^{-i\omega_2 x_0}}{N_2}-\frac{e^{-iW_2 x_0}}{\mathcal{N}_2}\right] \right]e^{i\vec p\cdot\vec x}\,.
\end{align}
Now we proceed with $x_0<0$
and compute
\begin{align}
S_F(x)&=S^{(<)}_F(x)\equiv - \bra{0} \bar\psi(0)\psi(x)\ket{0}     \,.
\end{align}
After some work similar to the one above, we find
\begin{align}\label{S<}
S_F^{(<)}(x)&=(i\slashed{\partial}+m+g_2\gamma_0\gamma_5\partial_0^2)
\int\frac{d^3\vec p}{(2\pi)^3}\left[  \frac{1}{2}  \left(\mathbb{1}_{4}-Q \right) \right. \notag
\\ &\times \left.   \left[\frac{ e^{i\omega_1 x_0}}{N_1}-\frac{ e^{iW_1 x_0}}{\mathcal{N}_1}
\right]+ \frac{1}{2}  \left(\mathbb{1}_{4}+Q \right) \right.\nonumber \\
 & \left. \left[\frac{
 e^{i\omega_2 x_0}}{N_2}-\frac{
 e^{iW_2 x_0}}{\mathcal{N}_2}\right] \right]e^{i\vec p\cdot\vec x}\,.
\end{align}

\begin{figure}
    \centering
    \includegraphics[scale=0.55]{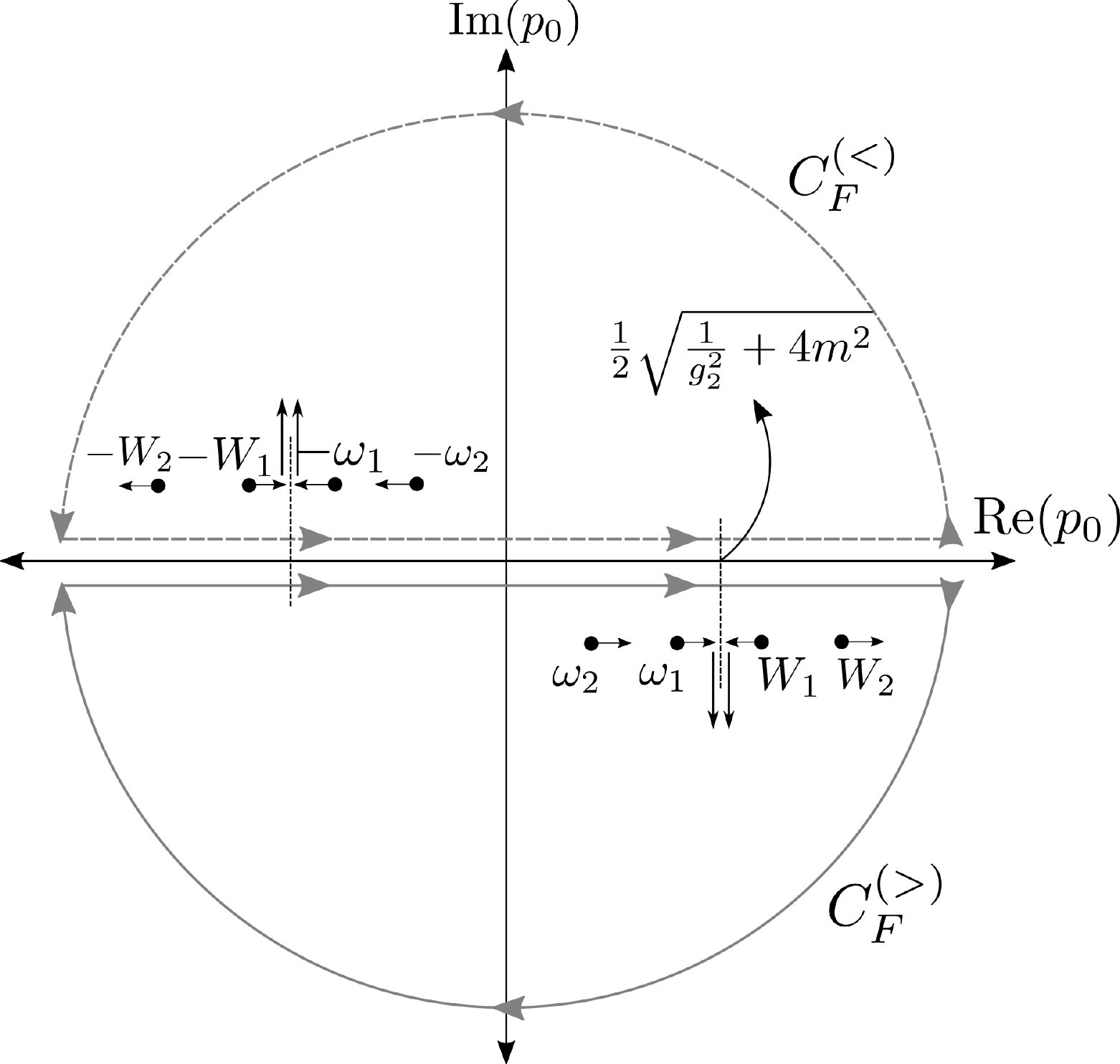}
    \caption{ \label{Fig1} The contour $C_F$ encloses the poles 
    $\omega_1,  \omega_2, W_1,W_2 $ in the lower half plane while it encloses the
     poles $-\omega_1,-  \omega_2, -W_1,-W_2 $
    in the upper half plane. At momentum
    $\vert\vec p\vert_{\text{max}}=\frac{1-4g_2^2m^2}{4g_2}$, the two poles 
    $\omega_1$ and $W_1$ have the same value
and from then  both move downwards parallel to the imaginary axis
as momentum increases. The two poles $\omega_2$ and $W_2$ go to 
infinity as the momentum increases and all the opposite sign poles have a similar behaviour.
}
\end{figure}

We are interested on
making contact with the four dimensional representation of the propagator 
with the pole prescription. Recall the inverse of the operator in the equation
 of motion~\eqref{timelike_spinor}
\begin{align}\label{Inv_M}
M^{-1}&=\frac{i\bar{M}N\bar{N}}{g_2^4(p_0^2-\omega_1^2)(p_0^2-W_1^2)
(p_0^2-\omega_2^2)(p_0^2-W_2^2)} \,,
\end{align}
with 
\begin{align}
\bar{M}N\bar{N}
&=(\slashed{p}+m-g_2p_0^2\gamma_0\gamma_5) \notag 
\\ &\times(p^2-m^2-g_2^2p_0^4+2g_2p_0^2p_i\gamma^i\gamma_0\gamma_5)\,.
\end{align}
In order to find the four dimensional representation of the propagator we need
the $i\epsilon$ prescription in the denominator of~\eqref{Inv_M} or
to define the Feynman contour $C_F$. 
We select a prescription for the propagator based on the contour $C_F$, see Fig~\eqref{Fig1}. 

Hence, let us write the Feynman propagator as
\begin{eqnarray}
S_F(x)=\int_{C_F}\frac{d^4p}{(2\pi)^4}S_F(p)e^{-ip\cdot x}\,,
\end{eqnarray}
with 
\begin{align}\label{S_Fprop}
S_F(p)&=\frac{i\bar{M}N\bar{N}}{ 	 \Lambda_+^2
(p+i\epsilon)  \Lambda_-^2 (p+i\epsilon)  } \,,
\end{align}
where from the expressions~\eqref{lambdas},
we are defining
\begin{align}
	 \Lambda_+^2(p+i\epsilon)&=-g_2^2(p_0+\omega_1-i\varepsilon) (p_0-\omega_1+i\varepsilon)   \notag \\ \times& 
(p_0+W_1-i\varepsilon) (p_0-W_1+i\varepsilon)  \,,
\notag \\
  \Lambda_-^2 (p+i\epsilon)&=-g_2^2 (p_0+\omega_2-i\varepsilon) (p_0-\omega_2+i\varepsilon)   
     \notag \\ \times&    (p_0+W_2-i\varepsilon) (p_0-W_2+i\varepsilon)   \,.
\end{align}
%...............................Fig1.....................................
%....................................................................
To compare with the previous calculation, let us consider
 $x_0>0$ and close the contour from below with the curve $C_F^{>}$, see Fig~\eqref{Fig1} to obtain
\begin{align}
S_F(x)&=\int_{C_F^{>}} \frac{dp_0}{(2\pi)}\int\frac{d^3\vec p}{(2\pi)^3}S_F(p)e^{-ip_0x_0+i\vec p\cdot\vec x} \,.
\end{align}
Integrating in $p_0$ produces 
\begin{align}
S_F(x)&=-\frac{(2\pi i)}{2\pi}\int\frac{d^3\vec p}{(2\pi)^3}  \sum_{i=1}^4 \left( 
 \text{Res}\left(S_F(p)e^{-ip_0x_0},q_i \right)  \right)\notag \\ 
&\times e^{i\vec p\cdot\vec x}\,,
\end{align}
where the sum runs over the
residues at the poles $q_1=\omega_1, q_2=\omega_2, q_3=W_1, q_4=W_2$ and $i=1,\dots 4$.

The evaluation of the residues are
\begin{align}
\text{Res}\left(S_F(p)e^{-ip_0x_0},\omega_1\right)&=\frac{i(\bar{M}N\bar{N})_{p_0=\omega_1}}
{g_2^2(\omega_1^2-\omega_2^2)(W_2^2-\omega_1^2)} \notag \\ &\times \frac{e^{-i\omega_1x_0}}{N_1} \,,
\end{align}
\begin{align}
\text{Res}\left(S_F(p)e^{-ip_0x_0},\omega_2\right)&=-\frac{i(\bar{M}N\bar{N})_{p_0
=\omega_2}}{g_2^2(\omega_1^2-\omega_2^2)(W_1^2-\omega_2^2)}  
 \notag \\ &\times   \frac{e^{-i\omega_2x_0}}{N_2} \,,
\end{align}
\begin{align}
\text{Res}\left(S_F(p)e^{-ip_0x_0},W_1\right)&=-\frac{i(\bar{M}N\bar{N})_{p_0=W_1}}
{g_2^2(W_1^2-\omega_2^2)(W_2^2-W_1^2)}\notag \\ &\times   \frac{e^{-iW_1x_0}}{\mathcal{N}_1} \,,
\end{align}
and
\begin{align}
\text{Res}\left(S_F(p)e^{-ip_0x_0},W_2\right)&=\frac{i(\bar{M}N\bar{N})_{p_0=W_2}}{g_2^2(W_2^2-\omega_1^2)(W_2^2-W_1^2)}   \notag
 \\ &\times  \frac{e^{-iW_2x_0}}{\mathcal{N}_2}  \,.
\end{align}
Considering the identities
\begin{align}
(\bar{M}N\bar{N})_{p_0=\omega_1}&=(4g_2\omega_1^2\vert\vec p\vert)  
(\omega_1\gamma_0+p_i\gamma^i+m-g_2\omega_1^2\gamma_0\gamma_5) \notag 
\\ &\times \frac{1}{2}  \left(\mathbb{1}_{4}-Q \right)  \,,
\end{align}
\begin{align}
(\bar{M}N\bar{N})_{p_0=\omega_2}&=(-4g_2\omega_2^2\vert\vec p\vert)
(\omega_2\gamma_0+p_i\gamma^i+m-g_2\omega_2^2\gamma_0\gamma_5)
\notag 
\\ &\times   \frac{1}{2}  \left(\mathbb{1}_{4}+Q \right) \,,
\end{align}
\begin{align}
(\bar{M}N\bar{N})_{p_0=W_1}&=(4g_2W_1^2\vert\vec p\vert)(W_1\gamma_0
+p_i\gamma^i+m-g_2W_1^2\gamma_0\gamma_5)
\notag 
\\ &\times \frac{1}{2}  \left(\mathbb{1}_{4}-Q \right) \,,
\end{align}
\begin{align}
(\bar{M}N\bar{N})_{p_0=W_2}&=(-4g_2W_2^2\vert\vec p\vert)(W_2\gamma_0
+p_i\gamma^i+m-g_2W_2^2\gamma_0\gamma_5)
\notag 
\\ &\times \frac{1}{2}  \left(\mathbb{1}_{4}-Q \right) \,,
\end{align}
and using the identities
\begin{align}
g_2^2(\omega_1^2-\omega_2^2)(W_2^2-\omega_1^2)&=4g_2\omega_1^2\vert\vec p\vert \,,   \notag    \\
g_2^2(\omega_1^2-\omega_2^2)(W_1^2-\omega_2^2)&= 4g_2\omega_2^2\vert\vec p\vert  \,,  \notag  \\
g_2^2(W_1^2-\omega_2^2)(W_2^2-W_1^2)&=4g_2W_1^2\vert\vec p\vert \,,   \notag    \\
g_2^2(W_2^2-\omega_1^2)(W_2^2-W_1^2)&=4g_2W_2^2\vert\vec p\vert  \,,
\end{align}
we can verify
\begin{align}
S_F(x)
&=\int \frac{d^3\vec p}{(2\pi)^3}
\left[(\omega_1\gamma_0+p_i\gamma^i+m-g_2\omega_1^2\gamma_0\gamma_5) \right. 
\notag \\ &\times  \left.  \frac{1}{2}  \left(\mathbb{1}_{4}-Q \right)   \frac{e^{-i\omega_1x_0}}{N_1}
 \right. \nonumber \\
&+(\omega_2\gamma_0+p_i\gamma^i+m-g_2\omega_2^2\gamma_0\gamma_5) \frac{1}{2} 
 \left(\mathbb{1}_{4}+Q \right)   \frac{e^{-i\omega_2x_0}}{N_2} \notag \\
&-(W_1\gamma_0+p_i\gamma^i+m-g_2W_1^2\gamma_0\gamma_5) \frac{1}{2} 
 \left(\mathbb{1}_{4}-Q \right) 
  \frac{e^{-iW_1x_0}}{\mathcal{N}_1} \notag \\
&\left.-(W_2\gamma_0+p_i\gamma^i+m-g_2W_2^2\gamma_0\gamma_5)  \frac{1}{2} 
 \left(\mathbb{1}_{4}+Q \right) \right. \notag \\ \times&   \left. 
 \frac{e^{-iW_2x_0}}{\mathcal{N}_2}   \right] \,.
\end{align}
Factorizing a global operator we arrive at
\begin{align}
S_F(x)&=(i\slashed{\partial}+m+g_2\gamma_0\gamma_5\partial_0^2)  \notag \\ &\times \int\frac{d^3\vec p}
{(2\pi)^3}\left[   \frac{1}{2}  \left(\mathbb{1}_{4}-Q \right)     \left[\frac{ e^{-i\omega_1 x_0}}{N_1}-
\frac{ e^{-iW_1 x_0}}{\mathcal{N}_1}\right]\right.\nonumber \\
 &\left.+  \frac{1}{2}  \left(\mathbb{1}_{4}+Q \right)     \left[\frac{
 e^{-i\omega_2 x_0}}{N_2}-\frac{
 e^{-iW_2 x_0}}{\mathcal{N}_2}\right] \right]e^{i\vec p\cdot\vec x}\,.
\end{align}
By comparing we arrive at the same result than the one obtained from the definition Eq.~\eqref{S>}.

Now we consider $x_0<0$ we close the contour in the upper half plane
\begin{align}
S_F(x)&=\int_{C_F^{<}} \frac{dp_0}{(2\pi)}\int\frac{d^3\vec p}{(2\pi)^3}S_F(p)e^{-ip_0x_0+i\vec p\cdot\vec x} \nonumber \\
&=\frac{(2\pi i)}{2\pi}\int\frac{d^3\vec p}{(2\pi)^3} \sum_{i=5}^8 \left(  \text{Res}\left(S_F(p)e^{-ip_0x_0},q_i \right)   \right)e^{i\vec p\cdot\vec x}
\end{align}
where now $q_5=-\omega_1, q_6=-\omega_2, q_7=-W_1, q_8=-W_2$ and $i=5,\dots 8$.

We have
\begin{align}
\text{Res}\left(S_F(p)e^{-ip_0x_0},-\omega_1\right)&=-\frac{i(\bar{M}N\bar{N})
_{p_0=\omega_1}}{g_2^2(\omega_1^2-\omega_2^2)(W_2^2-\omega_1^2)} 
 \notag \\ &\times  \frac{e^{i\omega_1x_0}}{N_1} \,,
\end{align}
\begin{align}
\text{Res}\left(S_F(p)e^{-ip_0x_0},-\omega_2\right)&=\frac{i(\bar{M}N\bar{N})
_{p_0=\omega_2}}{g_2^2(\omega_1^2-\omega_2^2)(W_1^2-\omega_2^2)}
 \notag \\ &\times \frac{e^{i\omega_2x_0}}{N_2} \,,
\end{align}
\begin{align}
\text{Res}\left(S_F(p)e^{-ip_0x_0},-W_1\right)&=\frac{i(\bar{M}N\bar{N})
_{p_0=W_1}}{g_2^2(W_1^2-\omega_2^2)(W_2^2-W_1^2)}
 \notag \\ &\times \frac{e^{iW_1x_0}}{\mathcal{N}_1} \,,
\end{align}
\begin{align}
\text{Res}\left(S_F(p)e^{-ip_0x_0},-W_2\right)&=-\frac{i(\bar{M}N\bar{N})
_{p_0=W_2}}{g_2^2(W_2^2-\omega_1^2)(W_2^2-W_1^2)}
 \notag \\ &\times \frac{e^{iW_2x_0}}{\mathcal{N}_2}  \,.
\end{align}
Consider
\begin{align}
&(\bar{M}N\bar{N})_{p_0=-\omega_1}=(4g_2\omega_1^2\vert\vec p\vert)
   \notag \\ &\times  (-\omega_1\gamma_0+p_i\gamma^i+m-g_2\omega_1^2\gamma_0\gamma_5) \frac{1}{2}  \left(\mathbb{1}_{4}-Q \right)\,,
\end{align}
\begin{align}
&(\bar{M}N\bar{N})_{p_0=-\omega_2}=(-4g_2\omega_2^2\vert\vec p\vert)
\notag \\ &\times   (-\omega_2\gamma_0+p_i\gamma^i+m-g_2\omega_2^2\gamma_0\gamma_5)  \frac{1}{2}  \left(\mathbb{1}_{4}+Q \right)\,,
\end{align}
\begin{align}
& (\bar{M}N\bar{N})_{p_0=-W_1}=(4g_2W_1^2\vert\vec p\vert)  \notag \\ &\times
(-W_1\gamma_0
+p_i\gamma^i+m-g_2W_1^2\gamma_0\gamma_5) \frac{1}{2}  \left(\mathbb{1}_{4}-Q \right)   \,,
\end{align}
\begin{align}
&(\bar{M}N\bar{N})_{p_0=-W_2}=(-4g_2W_2^2\vert\vec p\vert)
\notag \\ &\times  (-W_2\gamma_0+p_i\gamma^i+m-g_2W_2^2\gamma_0\gamma_5)  \frac{1}{2}  \left(\mathbb{1}_{4}+Q \right)\,.
\end{align}
We finally verify that
\begin{align}
S_F(x)&=\int \frac{d^3\vec p}{(2\pi)^3}\left[(-\omega_1\gamma_0+p_i\gamma^i+m-g_2
\omega_1^2\gamma_0\gamma_5) \right. \notag   \\  &   \left.  \frac{1}{2}  \left(\mathbb{1}_{4}-Q \right)   \frac{e^{i\omega_1x_0}}{N_1} 
+   (-\omega_2\gamma_0+p_i\gamma^i+m-g_2\omega_2^2\gamma_0\gamma_5)  \right. \nonumber \\
\times&   \frac{1}{2}  \left(\mathbb{1}_{4}+Q \right) \frac{e^{i\omega_2x_0}}{N_2} -  (-W_1\gamma_0+p_i\gamma^i+m-g_2W_1^2\gamma_0\gamma_5) \nonumber \\
&   \frac{1}{2}  \left(\mathbb{1}_{4}-Q \right)   \frac{e^{iW_1x_0}}{\mathcal{N}_1}   -(-W_2\gamma_0+p_i\gamma^i+m-g_2W_2^2\gamma_0\gamma_5) \nonumber \\
&\left.     \frac{1}{2}  \left(\mathbb{1}_{4}+Q \right) \frac{e^{iW_2x_0}}{\mathcal{N}_2}   \right] \,,
\end{align}
Again factorizing a global operators, we arrive at
\begin{align}
S_F(x)&=(i\slashed{\partial}+m+g_2\gamma_0\gamma_5\partial_0^2)\int\frac{d^3\vec p}{(2\pi)^3}
\left[\frac{1}{2}  \left(\mathbb{1}_{4}-Q \right)  \right. 
  \notag  \\ &\times  \left. \left[\frac{ e^{i\omega_1 x_0}}{N_1}-\frac{ e^{iW_1 x_0}}{\mathcal{N}_1}\right]\right.\nonumber \\
 &\left.+ \frac{1}{2}  \left(\mathbb{1}_{4}+Q \right)   \left[\frac{
 e^{i\omega_2 x_0}}{N_2}-\frac{
 e^{iW_2 x_0}}{\mathcal{N}_2}\right] \right]e^{i\vec p\cdot\vec x}\,,
\end{align}
 which is the same as obtained in \eqref{S<} with the definition.
%...................................................................................
\section{Microcausality }\label{sectionIV}
%...................................................................................
In quantum mechanics the property of 
causality means that
local observables commute at causally disconnected regions.
In relativistic field theory this assumption called microcausality is translated into the condition
\begin{align}
\left[ O(x), O(x') \right]  =0\,,    \qquad   \text{for}\; (x-x')^2<0   \,.
\end{align}
For a fermion theory, since observables are constructed from 
bilinear forms, it is enough to impose
\begin{align}
iS(x-x')=\lbrace \psi(x),\bar\psi(x')\rbrace\,,    \quad   \text{for}\;  (x-x')^2<0 \,.
\end{align}
In the model we are studying we can identify two sources of 
possible microcausality violations. The first one is related to
the breaking of Lorentz symmetry where the notion of light cone 
losses some of its properties due to superluminal propagation. The 
second one involves 
an indefinite metric leading to acausal propagation that has been 
extensively discussed in the literature by Lee and Wick and also in 
posterior works. 

We begin the study of microcausality by considering 
the decomposition~\eqref{fields12}, we obtain 
\begin{align}
\lbrace \psi(x),\bar\psi(x')\rbrace&=\lbrace \psi_1(x),\bar\psi_1(x')\rbrace  \notag \\ & +\lbrace \psi_2(x),\bar\psi_2(x')\rbrace \,.
\end{align}
We compute first
\begin{align}
& \lbrace \psi_1(x),\bar\psi_1(x')\rbrace=\sum_{r,s=1,2}\int\frac{d^3\vec p}{(2\pi)^3
}\frac{d^3\vec k}{(2\pi)^3}\frac{1}{\sqrt{N_r\bar N_s}}   \notag  \\ &  \lbrace a_p^ru^r(p)
e^{-i\omega_r   x_0+i\vec p\cdot\vec x}    +b_p^{r\dagger}v^r(p)e^{i\omega_r  x_0-i\vec p\cdot\vec x}, 
a_k^{s\dagger}u^{s\dagger}(k)    \nonumber\\
&\times   \gamma_0e^{i\bar\omega_s  x_0'-i\vec k\cdot\vec x'}  +b_k^{s}v^{s\dagger}(k)
\gamma_0e^{-i\bar\omega_s x_0'+i\vec k\cdot\vec x'}\rbrace \,.
\end{align}
We use the algebra~\eqref{Alg_pos} and the outer relations in~\eqref{uexp11} and~\eqref{uexp2}
to arrive at
\begin{align}		
& \lbrace \psi_1(x),\bar\psi_1(x')\rbrace=\int\frac{d^3\vec p}{(2\pi)^3}\left[\frac{1}{N_1}
\left( (\gamma_0\omega_1+\gamma^ip_i+m        \right.  \right.  \nonumber    \\    &  \left.   \left.  
      -g_2\omega_1^2\gamma_0\gamma_5)      \gamma_0     \frac 12 \left(\mathbb{1}_{4}-Q \right)  
   \gamma_0    e^{-i\omega_1(x_0-x_0')     }\right.\right.     \nonumber    \\
&\left.+(\gamma_0\omega_1-\gamma^ip_i-m+g_2\omega_1^2\gamma_0
\gamma_5)             \gamma_0 \frac 12  \left(\mathbb{1}_{4}-Q \right) 
   \gamma_0e^{i\omega_1(x_0-x_0')} \right)\nonumber \\
&+\frac{1}{N_2}\left( (\gamma_0\omega_2+\gamma^ip_i+m-g_2\omega_2^2\gamma_0\gamma_5)      
   \right.  \nonumber      \\    &  \left.         \gamma_0   \frac 12 
\left(\mathbb{1}_{4}+Q \right)   \gamma_0e^{-i\omega_2(x_0-x_0')}\right.\nonumber\\
& \left.\left.+(\gamma_0\omega_2-\gamma^ip_i-m+g_2\omega_2^2\gamma_0\gamma_5)  
  \right.  \right. \nonumber     \\    &  \left.   \left. 
\gamma_0   \frac 12     \left(\mathbb{1}_{4}+Q \right)     \gamma_0e^{i\omega_2(x_0-x_0')} \right) 
  \right]e^{i\vec p\cdot(\vec x-\vec x')}\,.
\end{align}
Taking $x'=0$ we get
\begin{align}
&\lbrace \psi_1(x),\bar\psi_1(0)\rbrace=\int\frac{d^3\vec p}{(2\pi)^3}\left[\frac{1}{N_1}
\left( (\gamma_0\omega_1+\gamma^ip_i+m-g_2\omega_1^2\gamma_0\gamma_5)        \right. \right.  \nonumber      \\    &  \left.  \left.    
  \frac 12   \left(\mathbb{1}_{4}-Q \right)    e^{-i\omega_1 x_0}      +(\gamma_0\omega_1-\gamma^ip_i-m+g_2\omega_1^2\gamma_0\gamma_5)      \right.\right.\nonumber\\
& \left.   
   \frac 12   \left(\mathbb{1}_{4}-Q \right)    e^{i\omega_1 x_0} \right)   \nonumber \\
& +\frac{1}{N_2}\left( (\gamma_0\omega_2+\gamma^ip_i+m-g_2\omega_2^2\gamma_0\gamma_5)      \right.  \nonumber      \\    &  \left.    
 \frac 12    \left(\mathbb{1}_{4}+Q \right)     e^{-i\omega_2 x_0}    +(\gamma_0\omega_2-\gamma^ip_i-m+g_2\omega_2^2\gamma_0\gamma_5)       \right.\nonumber\\
&\left.\left.  
    \frac 12   \left(\mathbb{1}_{4}+Q \right)   e^{i\omega_2 x_0} \right)   \right]e^{i\vec p\cdot\vec x}\,,
\end{align}		
and hence		
\begin{align}		
& \lbrace \psi_1(x),\bar\psi_1(0)\rbrace=  (i\slashed{\partial}+m+g_2\partial_0^{2}   \gamma_0\gamma_5)     \nonumber      \\    &     
\int\frac{d^3\vec p}{(2\pi)^3}\left[\frac{1}{N_1}\left(e^{-i\omega_1x_0}- e^{i\omega_1x_0}\right) \frac 12  \left(\mathbb{1}_{4}-Q \right)    \right.\nonumber\\
&\left.+\frac{1}{N_2}\left(e^{-i\omega_2x_0}-e^{i\omega_2x_0}  \right) \frac 12   \left(\mathbb{1}_{4}+Q \right)     \right]e^{i\vec p\cdot\vec x}\,.
\end{align}
Similar calculations lead to
\begin{align}
& \lbrace \psi_2(x),\bar\psi_2(0)\rbrace=(-1)(i\slashed{\partial}+
m+g_2\partial_0^{2}\gamma_0\gamma_5)      \nonumber      \\    &     \int\frac{d^3\vec p}{(2\pi)^3}\left[\frac{1}{\mathcal{N}_1}
\left(e^{-iW_1x_0}- e^{iW_1x_0}\right     )  \frac 12   \left(\mathbb{1}_{4}-Q \right)    \right.\nonumber\\
&\left.+\frac{1}{\mathcal{N}_2}\left(e^{-iW_2x_0}-e^{iW_2x_0}  \right)  \frac 12  \left(\mathbb{1}_{4}+Q \right)   \right]e^{i\vec p\cdot\vec x}\,.
\end{align}
We have the four dimensional representation of the anticommutator $\lbrace \psi(x),\bar\psi(x')\rbrace  $ 
by using the curve $C$ which encloses 
the eight poles. From~\eqref{Fig1}, where $C=C_F^{<}- C_F^{>}$ we can write
\begin{align}
S(x)= \hat {\bar M}   \hat N  \hat {\bar N}  \int_{C}\frac{d^4p}{(2\pi)^4}  \frac{e^{-ip\cdot x}}{ 	 \Lambda_+^2
(p+i\epsilon)  \Lambda_-^2 (p+i\epsilon)  } \,,
\end{align}
where
 \begin{align}
	 \hat  {\bar M}&=i\slashed{\partial}+m+g_2  \partial_0^{2}    \gamma_0  \gamma_5 \,, 
\notag   \\	
	\hat { {  N}}&=i\slashed{\partial}+m-g_2  \partial_0^{2} \gamma_0\gamma_5    \notag   \,,\\
	 \hat { \bar N}&=i\slashed{\partial}-m-g_2 \partial_0^{2} \gamma_0\gamma_5  \,.
\end{align}
We can always perform an observer transformation when both points are spacelike separated, 
 leaving us with $x=(0,\vec x)$. In this way we can integrate and obtain an integral proportional to
\begin{align}
& \int _C \frac{dp_0}{  (p_0^2-\omega_1^2)  (p_0^2-\omega_2^2)  (p_0^2-W_1^2)    (p_0^2-W_2^2)     }
\notag \\ &=    2\pi i   \left[  \frac{1}{   2 \omega_1 (\omega_1^2-\omega_2^2) (\omega_1^2-W_1^2) (\omega_1^2-W_2^2)   }\notag    \right. \\ &   \left. -\frac{1}
 { 2 \omega_1 (\omega_1^2-\omega_2^2) (\omega_1^2-W_1^2) (\omega_1^2-W_2^2)   }   \notag   \right. \\ &  \left.  +\frac{1}
 {  2 \omega_2 (\omega_2^2-\omega_1^2) (\omega_2^2-W_1^2) (\omega_2^2-W_2^2)  } \notag   \right.    \\ &  \left. -
\frac{1}{2 \omega_2 (\omega_2^2-\omega_1^2) (\omega_2^2-W_1^2) (\omega_2^2-W_2^2)  }    \notag   \right.
\\ &  \left. +\frac{1}
 {  2 W_1 (W_1^2-\omega_1^2) (W_1^2-\omega_2^2) (W_1^2-W_2^2)  }\notag   \right. \\ &  \left. -
\frac{1}{2 W_1   (W_1^2-\omega_1^2) (W_1^2-\omega_2^2) (W_1^2-W_2^2)   }   \notag  \right.   \\ &  \left.  +\frac{1}
 {  2 W_2 (W_2^2-\omega_1^2) (W_2^2-\omega_2^2) (W_2^2-W_1^2)  }\notag   \right.   \\ &  \left. -
\frac{1}{2 W_2 (W_2^2-\omega_1^2) (W_2^2-\omega_2^2) (W_2^2-W_1^2)  }   \notag   \right]   \\&    =0\,.
\end{align}
 The combination is always zero even when the poles $\omega_1$ and $W_1$ 
 become complex as can be seen in Fig.\eqref{Fig1}.
and therefore microcausality is preserved.
%................................................................................
\section{Tree-level unitarity}\label{sectionV}
%................................................................................
Recapitulating, we have found $\eta_{2,s}$
the metric associated to the indefinite Fock space
 which is not positive defined and will produce negative-norm states for
  odd occupation number of particles.
Generally, an indefinite metric $\eta$ can lead to 
a pseudo-unitary relation for the $S$-matrix 
\begin{align}
S^{\dag}\eta S=\eta\,,
\end{align}
which is not satisfactory to describe probability amplitudes.
 However, as was shown by Lee and Wick an indefinite-metric theory can have a chance
 to develop a fully unitary $S$-matrix.
In particular, they showed that by restricting 
the asymptotic space to contain only particles with positive-metric, it is 
possible to have a unitary condition for the $S$-matrix~\cite{LW,Boulware_Gross}.
%...............................Fig2.....................................
\begin{figure}
    \centering
    \includegraphics[scale=0.50]{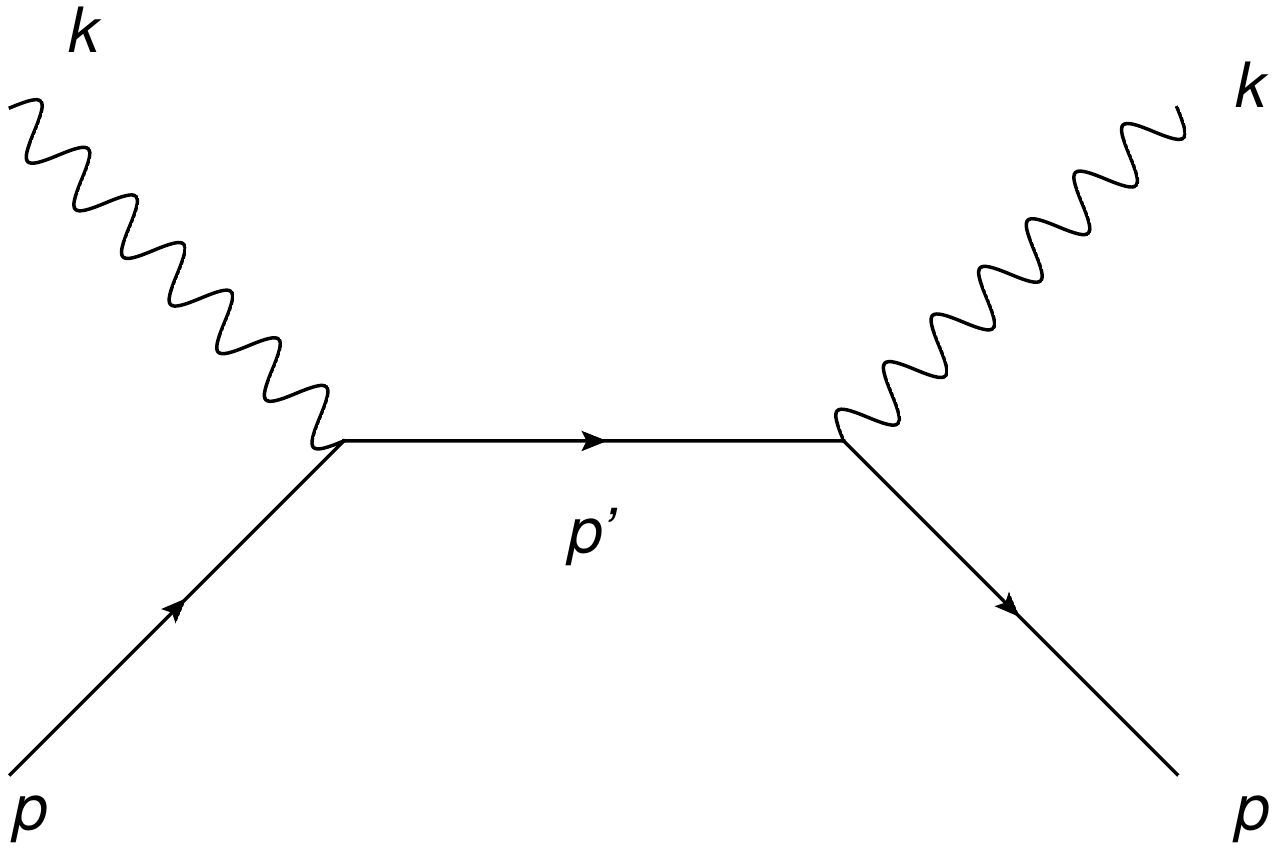}
    \caption{ \label{Fig2} The Compton scattering diagram in the analysis of tree-level order unitarity.
}
\end{figure}
%....................................................................

To study unitarity at tree-level we will use the tool of the
optical theorem and adopt the Lee-Wick prescription.
The optical theorem provides an important constraint equation
to test perturbative unitarity based on individual 
diagrams, which is well suited for our analysis.
Moreover, adopting the the Lee-Wick prescription in our model means 
that ghost states are unstable, and so, they will not appear in
external legs in any Feynman diagram. However, internal fermion lines propagating ghosts modes
are perfectly acceptable, leading to possible violations of unitarity. 
 Therefore to test these possible sources of unitarity violation, we focus our analysis on the class of
 diagrams describing $2\to 2$ processes at tree level with an 
 internal fermion line. 

Recall, the optical theorem has a simple expression
\begin{align}
2\text {Im }(M_{ii})=\sum_m \int d\Pi _m |M_{im}|^2\,,
\end{align}
where $M_{ii}$ is the amplitude for a forward scattering process. The sum runs over all possible
intermediate states and the integral 
over the phase space $d\Pi_m$ is restricted by momentum conservation.

We study the process of Compton scattering of electrons and positrons. We consider the incoming 
fermion or anti-fermion of spin $r$ to have momentum $p$ and the photon to have
 momentum $k$. The final state are another photon-electron or positron-electron pairs, as shown in Fig.~\eqref{Fig2}. 
 
We begin with the process involving the electron and 
denote the process  by
$e^{-}(p)\gamma(k)\to e^-(p)\gamma(k)$. According to the standard Feynman rules  
 the matrix element $\mathcal M\equiv \mathcal M (e^-\gamma \to e^-\gamma)$
 can be written as
\begin{align}\label{Amp_M}
\mathcal M&=  (-ie)^2 \int \frac{d^4 p' }{(2\pi)^4}\times (2\pi)^4  \delta^{(4)} 
(p+k-p') \notag  \\   &\times   {    \overline U}^{r,\lambda}(p,k)  S_{\text F}(p')  U^{r,\lambda}(p,k) \,,
\end{align}
where
\begin{align}
{ \overline U } ^{r,\lambda}(p,k)=  N^r_{p} N_{k}   \bar u^{r}(p) \varepsilon^{*(\lambda)}_{\mu}(k) \gamma^{\mu}  \,, \notag
\\
U^{r,\lambda}(p,k)=  N^r_{p} N_{k}  \gamma^{\mu}  u^{r}(p)   \varepsilon^{(\lambda)}_{\mu}(k)   \,,
\end{align}
and $N_{k}=\sqrt{ \frac{1}{2\omega_k} }$, with $\omega_k=|\vec k|$ is the usual photon 
normalization, $N^r_{p}=\sqrt{\frac{1}{N_r}}$ are the normalization constants
of Eqs.~\eqref{Norm_part} and the modified fermion propagator $S_F$ is given in Eq.~\eqref{S_Fprop}.

To compute the imaginary part we 
consider the decomposition in the propagator
\begin{align}
&\frac{1}{(p'_0-\Omega+i\epsilon) (p'_0+\Omega-i\epsilon)   } \notag \\ &=   \frac{1}{2\Omega} 
   \left[   \frac{1}{(p'_0-\Omega+i\epsilon)   } +
  \frac{1}{(p'_0+\Omega-i\epsilon)   }     \right] \,,
\end{align}
and use the identity
\begin{align}
\frac{1}{p'_0-\Omega+i\epsilon}=\mathcal P\frac{1}{p'_0-\Omega}-i\pi \delta(p'_0-\Omega)\,,
\end{align}
where $\mathcal P$ is the principal value.

Now, focusing on~\eqref{Amp_M}, we obtain
\begin{align}
&2 \text {Im}(\mathcal M)=   (2\pi)  e^2 \int \frac{d^3\vec  p' }{2N_r\omega_k}  \delta^{(4)} 
(p+k-p')  \notag \\ &\times  \bar u^{r}(p) \varepsilon^{*(\lambda)}_{\mu}(k) \gamma^{\mu}  
  \notag  \\   &\times  \sum_{s=1,2}  \left(  \frac {   \bar M' N' {\bar N }'  }{2\omega'_s g_2^4 (\omega_s^{\prime  2}-\omega_2^{\prime 2})
 (\omega_s^{\prime  2}-W_1^{\prime 2})(\omega_s^{\prime  2}-W_2^{\prime 2})} \right)_{p_0'=\omega'_s} 
   \notag \\ &\times   \gamma^{\mu}  u^{r}(p)   \varepsilon^{(\lambda)}_{\mu}(k)  \,,
\end{align}
where the prime remind us that it is evaluated in $p_s '=(\omega_s(\vec p ^{\prime}),\vec p ')$. Note that the ghost 
states do not appear in the sum since by momentum conservation their contribution vanishes
when going on-shell.

Now, we will relate the amplitude with 
the total cross section $\sigma$ of the process $ e^-\gamma \to  e^-$.
We denote the total cross section by $\widehat {\mathcal M}\equiv \mathcal M (e^-\gamma \to e^-)$ and write
\begin{align}
\sigma=  \sum_{s=1,2}  \int \frac{d^3 \vec p'  }{(2\pi )^3 } \times (2\pi)^4\delta^{(4)} 
(p+k-p') |  \widehat {\mathcal M}_s |^2  \,.
\end{align}
with 
\begin{align}
 \widehat {\mathcal M}_s  = ie \frac{1}{\sqrt{N'_s}}  \frac{1}{\sqrt{N_r}}  \frac{1}{\sqrt{2\omega_k}}  
   {\bar u}^s (p') \gamma^{\nu}   u^r (p)  \varepsilon^{(\lambda)}_{\nu}(k)     \,.
\end{align}
The integral in phase space selects only particles which have the chance to satisfy momentum conservation.
We arrive at
\begin{align}
\sigma&= (2\pi) \sum_{s=1,2}  \int \frac{d^3 \vec p'  }{ 2N_r\omega_k} \delta^{(4)} 
(p+k-p')  \notag \\&   (ie  \frac{1}{\sqrt{N'_s}}   {\bar u}^s (p') \gamma^{\nu}   u^r (p)
  \varepsilon^{(\lambda)}_{\nu}(k)  )  ^{\dag} \notag  \\ \times&
  (ie \frac{1}{\sqrt{N'_s}}  {\bar u}^s (p') \gamma^{\nu}   u^r (p)  \varepsilon^{(\lambda)}_{\nu}(k)  )      \,,
\end{align}
then
\begin{align}
\sigma&=  (2\pi) e^2   \int \frac{d^3 \vec p'  }{2N_r\omega_k} \delta^{(4)} 
(p+k-p')    \bar  u^r (p) \gamma^{\nu} \varepsilon^{* (\lambda)}_{\nu}(k)  \notag  \\ \times&
  \left[ \sum_{s=1,2}   \frac{{ u}^s (p')   {\bar u}^s (p')  }{N'_s}  \right]       \gamma^{\mu}   u^r (p)  \varepsilon^{(\lambda)}_{\mu}(k)      \,.
\end{align}
To connect with the left hand side, consider the relations
\begin{align}
u^{(1)} (p) \bar {u}^{(1)}(p)&=   \left( \frac{\bar M N\bar N }{2(p^2-m^2-g_2^2 p_0^4)}\right)_{p_0=\omega_1} \,,
\notag \\
u^{(2)} (p) \bar {u}^{(2)}(p)&=   \left( \frac{\bar M N\bar N }{2(p^2-m^2-g_2^2 p_0^4)}\right)_{p_0=\omega_2} \,,
\end{align}
and the identities
\begin{align}
2(p^2-m^2-g_2^2 p_0^4)_{p_0=\omega_1} &=-g_2^2(\omega_1^2-\omega_2^2)(\omega_1^2-W_2^2) \,,
  \nonumber \\ 
2(p^2-m^2-g_2^2 p_0^4)_{p_0=\omega_2} &=-g_2^2(\omega_2^2-\omega_1^2)(\omega_2^2-W_1^2)  \,.
\end{align}
Hence we can write 
\begin{align}
& \frac{ u^{(1)} (p') \bar {u}^{(1)}(p')}{N'_1}=  \\&   \left( \frac{\bar M' N'\bar N' }{2\omega'_1 g_2^4  
 (\omega_1^{\prime 2}-\omega_2^{\prime 2})
 (\omega_1^{\prime 2}-W_1^{\prime 2})(\omega_1^{\prime 2}-W_2^{\prime2})}\right)_{p'_0=\omega'_1}\notag  \,,
\end{align}
and
\begin{align}
&\frac{ u^{(2)} (p') \bar {u}^{(2)}(p')}{N'_2} =  \\&
  \left( \frac{\bar M' N'\bar N' }{2\omega'_2 g_2^4  
   (\omega_2^{\prime 2}-\omega_1^{ \prime 2})(\omega_2^{\prime 2}-W_1^{\prime 2})
   (\omega_2^{\prime 2}-W_2^{\prime 2})}\right)_{p_0=\omega_2} \notag \,,
\end{align}
Finally, we have
\begin{align}
& \sum_{s=1,2}   \frac{{ u}^s (p')   {\bar u}^s (p')  }{N_s}    \\&   =  \sum_{s=1,2}  \left(  
\frac {   \bar M' N' {\bar N }'  }{2\omega'_s g_2^4 (\omega_s^{\prime  2}-\omega_2^{\prime 2})
 (\omega_s^{\prime  2}-W_1^{\prime 2})(\omega_s^{\prime  2}-W_2^{\prime 2})} \right)_{p_0'=\omega'_s} \notag  \,.
\end{align}
In this way we can prove the identity and thereby the validity of the optical theorem 
showing that unitarity is preserved for these processes at tree-level. The Compton scattering of a positron
follows by similar arguments.
%................................................................................
\section{Final Remarks}\label{sectionVI}
%................................................................................
We have studied a modified QED model containing Lorentz-violating dimension-five operators of Myers-Pospelov type in the fermion sector. 
The effective model, also a subset of the nonminimal SME framework, introduces Lorentz violation through a four-vector $n$. 
We have set $n$ to be purely timelike with a resulting Lagrangian 
coupling the effective terms to higher-order time derivatives. We have quantized the nonminimal Lorentz-violating model and 
distinguished at each step in the calculations between the corrected particle fields versus the new degrees of freedom that enter through the higher-order operators.
We have identified the positive and negative metrics that characterize the indefinite Fock space and found that 
ghost states with odd occupation numbers have a negative norm.

 The charge conjugation even sector of higher-order modified fermions has been less 
 explored than the charge conjugation odd sector, making it an excellent arena to explore kinematic modifications. In particular, we have found that the theory doubles the usual number of spinors and energy solutions of the dispersion relation concerning the standard theory.
We have found that the Hamiltonian is stable and hermitian in the effective region, although it can develop complex eigenvalues for higher energies and lose its hermitian property.

The new pole structure is essential to construct the propagator and fix the prescription for the curve $C_F$ in the $p_0$-complex plane. We have seen that 
the poles related to negative energies $\omega_2, W_2$ remain in the real axis while the poles $\omega_1, W_1$ can move vertically in the imaginary axis for energies above
$\vert p_{\max}\vert=\frac{1-4g_2^2m^2}{g_2}$.
We have studied microcausality by focussing on an anticommutator between fields. We have found that microcausality 
can be preserved by considering the pole structure and its evolution properties in the complex $p_0$-plane.
We have considered the forward scattering process involving fermion (antifermion) and photon pairs 
with an internal fermion line to study unitarity. We have found that unitarity is preserved at tree level by 
applying the Lee-Wick prescription and using the optical theorem to test perturbative unitarity.
\medskip
%..................................................................................
\acknowledgments
%..................................................................................
JLS thanks the Spanish Ministery of Universities and the European Union  
Next Generation EU/PRTR for the funds through the Maria 
Zambrano grant to attract international talent 2021 program. CMR 
has been funded by Fondecyt Regular grant No. 1191553, Chile, and wants to thank
 the kind hospitality of 
JLS at the University of Barcelona, where this work was finished. CR acknowledges 
support by ANID fellowship No. 21211384 from 
the Government of Chile and Universidad de Concepci\'on.
\appendix
%................................................................................
\section{Modified kinematics}\label{App:A}
%................................................................................
Here we 
derive the spinor solutions of the equation of motion~\eqref{eqM} and \eqref{neg_sol}. We give 
various type of orthogonality and outer product relations satisfied by the spinors.
%................................................................................
\subsection{Spinor solutions}\label{subappendixA}
%................................................................................
We start with the set of equations~\eqref{posit_sol} 
and multiply the second equation by $	p_0-g_2p_0^2-(\vec{p}\cdot\vec{\sigma} )$ to obtain
\begin{align}\label{eigen1_eq}
m^2\chi_1  &=\left(	p_0-g_2p_0^2-(\vec{p}\cdot\vec{\sigma} ) \right) \nonumber \\ &\times
 (p_0+g_2p_0^2+(\vec{p}\cdot\vec{\sigma} ))   \chi_1 \,.
\end{align}
To solve this equation we introduce the the two bi-spinors $\xi^{(\pm)}(\vec p)$,
given by
\begin{align}
\xi^{(+)}(\vec p)&= \frac{1}{\sqrt{  2|\vec p| \left(|\vec p|+p^3 \right)}}   \left(\begin{array}{c}
|\vec p| +p^3 	\\
p^1+ip^2	
\end{array}\right) \,,
\\
\xi^{(-)}(\vec p)&= \frac{1}{\sqrt{  2|\vec p| (|\vec p|-p^3)}}   \left(\begin{array}{c}
p^1-ip^2		\\  |\vec p| -p^3
\end{array}\right) \,,
\end{align}
which satisfy the properties
\begin{align}\label{propc}
	 (\vec{p}\cdot\vec{\sigma}  )  \xi^{(\pm)}  (\vec p) &= |\vec p| \xi^{(\pm)}(\vec p) \,,
\\
	 (\vec{p}\cdot\vec{\sigma}  )  \xi^{(\pm)}  (-\vec p) &=- |\vec p| \xi^{(\pm)}(-\vec p) \,,
\end{align}
and the orthogonality relations 
\begin{align}\label{Ort_prop_bi}
\xi^{(+) \dag}(\vec p)  \xi^{(+)}(\vec p)&=\xi^{(-) \dag}(\vec p)  \xi^{(-)}(\vec p)=1\,,
\\
\xi^{(+) \dag }(\vec p)  \xi^{(-)}(-\vec p)&=\xi^{(-) \dag }(-\vec p)  \xi^{(+)}(\vec p)=0\,.
\end{align}
In addition, we list the relations 
\begin{eqnarray}\label{ep}
\xi^{(+)  }(\vec p)  \xi^{(+) \dag }(\vec p)&=&\xi^{(-)  }(\vec p)  \xi^{(-) \dag }(\vec p)\nonumber 
\\ &=&\frac{1}{2}\left(1+\frac{\vec \sigma \cdot \vec p} {|\vec p|} \right)\,,
\\
\xi^{(+)  }(-\vec p)  \xi^{(+) \dag }(-\vec p)&=&\xi^{(-)  }(-\vec p)  \xi^{(-) \dag }(-\vec p)
 \nonumber \\ &=&\frac{1}{2}\left(1-\frac{\vec \sigma \cdot \vec p} {|\vec p|} \right)\,.
\end{eqnarray}
Returning to our derivation, we select
 $\chi^{(+)}_1 ( \vec p)=A_1 \xi^{(+)}(\vec p)$ in Eq.~\eqref{eigen1_eq}  and using the property~\eqref{propc},
 it can be shown that the bi-spinor solves the equation of motion given that 
its momentum satisfies the dispersion relation $ \Lambda _+^2(p)=0$. 
 
 According to~\eqref{posit_sol}, we have
 $ \chi^{(+)}_2 (\vec p)=\frac{A_1}{m} 
(p_0+g_2p_0^2+(\vec{p}\cdot\vec{\sigma} ))  \xi^{(+)}(\vec p)$ which produces 
the two energy-dependent solutions 
\begin{align}
	u^{(1)}( p)&=A_1\left(\begin{array}{c}  \xi^{(+)}(\vec p)\\ 
	\left( \frac{p_0+g_2p_0^2+\vec{p}\cdot\vec{\sigma} }{m} \right) \xi^{(+)}(\vec p)
	\end{array}\right)_{p_0=\omega_1}  \,,
\end{align}
and
\begin{align}
	U^{(1)}( p)&=\mathcal{A}_1\left(\begin{array}{c}  \xi^{(+)}(\vec p)\\ 
	\left( \frac{p_0+g_2p_0^2+\vec{p}\cdot\vec{\sigma} }{m} \right) \xi^{(+)}(\vec p)
	\end{array}\right)_{p_0=W_1}  \,. 
\end{align}
In a similar fashion, let us choose a different  bi-spinor $\chi_1^{(-)} (\vec p)=A_2 \xi^{(-)}(-\vec p)$
with its momentum satisfying the dispersion relation $ \Lambda _-^2(p)=0$. The bi-spinor produces 
the two solutions
\begin{align}
	u^{(2)}( p)&=A_2\left(\begin{array}{c}  \xi^{(-)}(-\vec p)\\ 
	\left( \frac{p_0+g_2p_0^2+\vec{p}\cdot\vec{\sigma} }{m} \right) \xi^{(-)}(-\vec p)
	\end{array}\right)_{p_0=\omega_2}  \,,
\end{align}
and
\begin{align}
	U^{(2)}( p)&=\mathcal{A}_2\left(\begin{array}{c}  \xi^{(-)}(-\vec p)\\ 
	\left( \frac{p_0+g_2p_0^2+\vec{p}\cdot\vec{\sigma} }{m} \right) \xi^{(-)}(-\vec p) 
	\end{array}\right)_{p_0=W_2}  \,.
\end{align}
For positive-energy spinors associated to particle and ghost modes
we choose the normalization constants as
\begin{eqnarray}
A_1=\mathcal A_1&=& \sqrt{p_0-g_2p_0^2-|\vec{p}| }    \,,
\\
A_2=\mathcal A_2&=& \sqrt{p_0-g_2p_0^2+|\vec{p}| }    \,.
\end{eqnarray}
In this way we obtain the spinors given in~\eqref{spinors1} and~\eqref{spinors2}. 

Now we search for negative-energy solutions 
which satisfy the equation of motion~\eqref{neg_sol}.
We multiply the first equation in \eqref{bispinor2} by 
$	p_0-g_2p_0^2+(\vec{p}\cdot\vec{\sigma} ) $ and obtain
\begin{eqnarray}\label{eigen1_eq2}
m^2\phi_2&=& (	p_0-g_2p_0^2+(\vec{p}\cdot\vec{\sigma} ) ) \nonumber
\\ &\times& (p_0+g_2p_0^2-(\vec{p}\cdot\vec{\sigma} ))   \phi_2 \,. 
\end{eqnarray}
The equation can be satisfied by choosing 
$\phi_2  (\vec p)=B_1 \xi^{(-)}(-\vec p)$ with on-shell momentum satisfying $  \Lambda _+^2=0$. In a analogous form 
we have
\begin{equation}
	v^{(1)}(p)=B_1\left(\begin{array}{c}  -\left( \frac{p_0+g_2p_0^2
	-\vec{p}\cdot\vec{\sigma} }{m} \right) \xi^{(-)}(-\vec p)\\ 
	 \xi^{(-)}(-\vec p) \end{array}\right)_{p_0=\omega_1}  \,,
\end{equation}
and
\begin{equation}
	V^{(1)}( p)=\mathcal{B}_1 \left(\begin{array}{c} -\left( \frac{p_0+g_2p_0^2-\vec{p}\cdot\vec{\sigma} }{m} \right)  \xi^{(-)}(-\vec p)\\ 
	\xi^{(-)}(-\vec p) \end{array}\right)_{p_0= W_1}  \,.
\end{equation}
Now, we choose
$\phi_2 (\vec p)=B_2 \xi^{(+)}(\vec p)$ in~\eqref{eigen1_eq2},
with momentum solving $  \Lambda _-^2=0$, which produces 
the two spinor solutions
\begin{equation}
	v^{(2)}( p)=B_2 \left(\begin{array}{c} -\left( \frac{p_0+g_2p_0^2-\vec{p}\cdot\vec{\sigma} }{m} \right) \xi^{(+)}(\vec p)\\ 
	 \xi^{(+)}(\vec p) \end{array}\right)_{p_0= \omega_2}  \,,
\end{equation}
and
\begin{equation}
	V^{(2)}( p)=\mathcal{B}_2  \left(\begin{array}{c} -\left( \frac{p_0+g_2p_0^2-\vec{p}\cdot\vec{\sigma} }{m} \right)  \xi^{(+)}(\vec p)\\ 
	\xi^{(+)}(\vec p) \end{array}\right)_{p_0= W_2}  \,.
\end{equation}
For this set of negative-energy spinors, we choose the normalization constants to be
\begin{eqnarray}
B_1=\mathcal B_1&=&- \sqrt{p_0-g_2p_0^2-|\vec{p}| } \,, \\
B_2=\mathcal B_2&=& -\sqrt{p_0-g_2p_0^2+|\vec{p}| }    \,,
\end{eqnarray}
and we obtain the solutions~\eqref{v_1} and~\eqref{v_2}.
%........................................................................................
\subsection{Inner product relations }\label{subappendixB}
%........................................................................................
For the many expressions it is convenient to introduce the notation for the
positive-energy spinors as
\begin{align}
	u^{(1)}( p)&=\left(\begin{array}{c}  A \xi^{(+)}(\vec p)\\ 
	B \xi^{(+)}(\vec p) \end{array}\right)_{p_0=\omega_1}  \,,
\nonumber \\
	U^{(1)}( p)&=\left(\begin{array}{c} A   \xi^{(+)}(\vec p)\\ \label{spinors1}
	B  \xi^{(+)}(\vec p) \end{array}\right)_{p_0=W_1}  \,.
\end{align}
\begin{align}
	u^{(2)}( p)&=\left(\begin{array}{c} C  \xi^{(-)}(-\vec p)\\ 
	D  \xi^{(-)}(-\vec p) \end{array}\right)_{p_0=\omega_2}  \,,
\nonumber \\
	U^{(2)}( p)&=\left(\begin{array}{c}  C \xi^{(-)}(-\vec p)\\ \label{spinors2}
	D \xi^{(-)}(-\vec p) \end{array}\right)_{p_0=W_2}  \,.
\end{align}
and also the negative-energy spinors
\begin{eqnarray}
	v^{(1)}( p)&=&\left(\begin{array}{c}  B \xi^{(-)}(-\vec p)\\ 
	 - A \xi^{(-)}(-\vec p) \end{array}\right)_{p_0= \omega_1}  \,,
\nonumber \\
	V^{(1)}( p)&=&\left(\begin{array}{c}  B  \xi^{(-)}(-\vec p)\\ \label{v_1}
	- A  \xi^{(-)}(-\vec p) \end{array}\right)_{p_0= W_1}  \,.
\end{eqnarray}
\begin{eqnarray}
	v^{(2)}(p)&=&\left(\begin{array}{c} D \xi^{(+)}(\vec p)\\ 
	- C \xi^{(+)}(\vec p) \end{array}\right)_{p_0= \omega_2}  \,,
\nonumber \\
	V^{(2)}( p)&=&\left(\begin{array}{c}  D \xi^{(+)}(\vec p)\\ \label{v_2}
-C \xi^{(+)}(\vec p) \end{array}\right)_{p_0= W_2}  \,, 
\end{eqnarray}
with
\begin{align}
	A&= \sqrt{p_0-g_2p_0^2-|\vec{p}| } \,,\\
	B&= \sqrt{p_0+g_2p_0^2+|\vec{p}| }\,, \\
	C&=\sqrt{p_0-g_2p_0^2+|\vec{p} |}\,, \\
	D&=\sqrt{p_0+g_2p_0^2-|\vec{p} |}\,.
\end{align}
In particular, with the property~\eqref{Ort_prop_bi} we find
\begin{eqnarray}
	u^{(1)}(p) u^{(1)\dag}(p)=(A^2+B^2)_{p_0=\omega_1}\,,
\end{eqnarray}
resulting in
\begin{eqnarray}
	u^{(1)\dag }(p) u^{(1)}(p)=2\omega_1\,.
\end{eqnarray}
The same occurs for $U^{(1)}(p)$ leading to the expressions in~\eqref{Inner_uv} and~\eqref{Inner_UV}.

Now consider
\begin{align}
	 {\bar u}^{(1)}(p) u^{(1)}(p)&=2(AB)_{p_0=\omega_1}=2m\,,  \\  {\bar v}^{(1)}(p) v^{(1)}(p)&=-2(AB)_{p_0=\omega_1}=-2m\,, 
\end{align}
and again we get the relations listed in~\eqref{Innerbar_uv} and ~\eqref{Innerbar_UV}.

Let us define the operators
\begin{align}
 q^{(+)}_{rs}(p)&=  \mathbb{1}_{4}- g_2(\omega_r+\omega_s)   \gamma_5 \,,
\end{align}
\begin{align}
 { q}^{(-)}_{rs}(p)&=  \mathbb{1}_{4}+  g_2(\omega_r+\omega_s)   \gamma_5 \,,
\end{align}
and
\begin{align}
 Q^{(+)}_{rs}(p)&=  \mathbb{1}_{4}- g_2(W_r+W_s)   \gamma_5 \,,
\end{align}
\begin{align}
 { Q}^{(-)}_{rs}(p)&=  \mathbb{1}_{4}+  g_2(W_r+W_s)   \gamma_5 \,,
\end{align}
where $\mathbb{1}_{4} $ is the unit $4\times 4$ matrix and $r,s=1,2$.

To prove the next relations we follow a trick. Consider the element
\begin{align}
& u^{r\dag}(p)\gamma_0   \left(  \gamma^ip_i-m   \right) u^s(p) \,,
\end{align}
which can be written using the equations of motion as
\begin{align}
&  u^{r\dagger}(p)\left( -\omega_s+g_2\gamma_5(\omega_s)^2 \right) u^s(p)   \,,
\end{align}
or
\begin{align}
  &u^{r\dagger}(p) \left(-\omega_r+g_2\gamma_5(\omega_r)^2\right)u^s(p) \,,
\end{align}
we arrive at 
\begin{align}
 u^{r\dagger}(p)  & \left((\omega_s-\omega_r)   -g_2\gamma_5((\omega_s)^2-(\omega_r)^2)  \right) \notag  \\ &\times u^s(p)=0 \,,
\end{align}
and in the case $\omega_r \neq \omega_s$, we have
\begin{align}
u^{r\dagger}(p) q^{(+)}_{rs}  u^s(p)=0\,.
\end{align}
We can write 
\begin{align}\label{G_prod}
 u^{(r)\dag}(\vec{p})  q^{(+)}_{rs} u^{(s)}(\vec{p})
 =C_r\delta^{rs}  \,,
\end{align}
where $C_r$ is a constant that has to be determined. Doing the same with all other contributions, and computing directly for the 
same energies, i.e., $\omega_r =\omega_s$, we find for particle spinors
\begin{align}\label{Prod_uu}
 u^{(1)\dag}(\vec{p})  q^{(+)}_{11} u^{(1)}(\vec{p})
 &=N_1  \,,
\notag \\
 u^{(2)\dag}(\vec{p}) q^{(+)}_{22}  u^{(2)}(\vec{p})
 &=N_2  \,,
\notag  \\
 v^{(1)\dag}(\vec{p}) { q}^{(-)}_{11}  v^{(1)}(\vec{p})
 &=N_1 \,,
\notag   \\
 v^{(2)\dag}(\vec{p}) { q}^{(-)}_{22}   v^{(2)}(\vec{p})
& =  N_2 \,,
\end{align}
and for ghost spinors
\begin{align}\label{Prod_UU}
 U^{(1)\dag}(\vec{p}) Q^{(+)}_{11} U^{(1)}(\vec{p})
& =-  \mathcal N_1  \,,
\notag \\
 U^{(2)\dag}(\vec{p}) Q^{(+)}_{22}  U^{(2)}(\vec{p})
& =- \mathcal N_2  \,,
\notag \\
 V^{(1)\dag}(\vec{p})  { Q}^{(-)}_{11}  V^{(1)}(\vec{p})
 &=      - \mathcal N_1  \,,
\notag \\
 V^{(2)\dag}(\vec{p})  { Q}^{(-)}_{22}  V^{(2)}(\vec{p})
& =- \mathcal N_2   \,.
\end{align}
We define positive normalization constants~\eqref{Norm_part}  and~\eqref{Norm_neg}  with respect to
those inner products,
where for negative-metric states we have taken the absolute value.

In the same way one can prove that for any $r,s$
one has the expressions
\begin{align}\label{Mixed}
u^{r\dagger}(p)  (1+g_2\gamma_5(\omega_s-\omega_r))  v^s(-p)&=0\,,
\notag \\
u^{r\dagger}(p)   (1-g_2\gamma_5(W_s+\omega_r)) U^s(p)&=0\,,
\notag \\
u^{r\dagger}(p)    (1+g_2\gamma_5(W_s-\omega_r))   V^s(-p)&=0\,,
\notag \\
U^{r\dagger}(p)    (1+g_2\gamma_5(\omega_s-W_r))   v^s(-p)&=0\,,
\notag \\
U^{r\dagger}(p)    (1+g_2\gamma_5(W_s-W_r))   V^s(-p)&=0 \,,
\notag \\
v^{r\dagger}(-p)   (1+g_2\gamma_5(W_s+\omega_r))   V^s(-p)&=0\,.
\end{align}
%.....................................................................
\subsection{Outer product relations}\label{subappendixC}
%.....................................................................
Here we prove outer product relations that are used for the quantization.
We start to consider
\begin{align}
 u^{(1)}  \bar {u}^{(1)} &=	\left(\begin{array}{c c}
m & (\omega_1-
g_2\omega_1^2-(\vec{p}\cdot\vec{\sigma} ) )   \notag  \\
 	(	\omega_1+g_2\omega_1^2+(\vec{p}\cdot\vec{\sigma} )  ) 
	 & m 
	\end{array}\right) \\ & \otimes \frac 12  (1+\frac{ \vec \sigma \cdot \vec p}{|\vec p|})  \,,
\end{align}
where we have used the property of the bi-spinors~\eqref{ep}.

Noting that 
\begin{align}\label{eq_motion2}
	\bar  M(\omega_1,\vec p)= \left(\begin{array}{c c}
		m& \omega_1-g_2\omega_1^2-(\vec{p}\cdot\vec{\sigma} )   \\
		\omega_1+g_2\omega_1^2+(\vec{p}\cdot\vec{\sigma} ) &m
	\end{array}\right)  \,.
\end{align}
and using~\eqref{Op_Mbar} we can write
\begin{align}\label{uexp11}
u^{(1)}(p)\bar{u}^{(1)}(p)&=(\gamma_0 \omega_1 +\gamma^ip_i
+m-g_2\omega_1^2\gamma_0\gamma_5)   
\notag \\     \times& \frac 12 \left(\mathbb{1}_{4} -Q \right)  \,.
\end{align}
\begin{align} \label{uexp2}
u^{(2)}(p)\bar{u}^{(2)}(p)&=(\gamma_0\omega_2+\gamma^ip_i+m-g_2\omega_2^2\gamma_0\gamma_5)\notag \\ \times&
 \frac 12 (\mathbb{1}_{4} +Q)\,,
\end{align}
\begin{align}
U^{(1)}(p)\bar{U}^{(1)}(p)&=(\gamma_0W_1+\gamma^ip_i+m-g_2W_1^2\gamma_0\gamma_5)\notag \\ \times&
 \frac 12 \left(\mathbb{1}_{4} -Q \right) \,,
\end{align}
\begin{align}
U^{(2)}(p)\bar{U}^{(2)}(p)&=(\gamma_0W_2+\gamma^ip_i+m-g_2W_2^2\gamma_0\gamma_5)\notag \\ \times&
 \frac 12 \left(\mathbb{1}_{4} +Q \right) \,,
\end{align}
\begin{align}
v^{(1)}(-p)\bar{v}^{(1)}(-p)&=(\gamma_0\omega_1-\gamma^ip_i-m+g_2\omega_1^2\gamma_0\gamma_5)\notag \\ \times&
 \frac 12 \left(\mathbb{1}_{4} -Q \right) \,,
\end{align}
\begin{align}
v^{(2)}(-p)\bar{v}^{(2)}(-p)&=(\gamma_0\omega_2-\gamma^ip_i-m+g_2\omega_2^2\gamma_0\gamma_5)\notag \\ \times&
 \frac 12 \left(\mathbb{1}_{4} +Q \right) \,,
\end{align}
\begin{align}
V^{(1)}(-p)\bar{V}^{(1)}(-p)&=(\gamma_0W_1-\gamma^ip_i-m+g_2W_1^2\gamma_0\gamma_5)\notag \\ \times&
 \frac 12 \left(\mathbb{1}_{4} -Q \right) \,,
\end{align}
\begin{align}
V^{(2)}(-p)\bar{V}^{(2)}(-p)&=(\gamma_0W_2-\gamma^ip_i-m+g_2W_2^2\gamma_0\gamma_5)\notag \\ \times&
 \frac 12 \left(\mathbb{1}_{4} +Q \right) \,,
\end{align}
where the operator $Q$ is defined in~\eqref{Qtime}.

Let us multiply the above identities by the left with $\gamma_0$, and add conveniently, we obtain
\begin{align}\label{uu1+vv1dag}
&u^{(1)}(p)u^{(1)\dagger}(p)+v^{(1)}(-p)v^{(1)\dagger}(-p) 
\notag \\  &=\omega_1    (\mathbb{1}_{4} -Q)\,,
 \end{align} 
\begin{align}\label{uu1-vv1dag}
   & u^{(1)}(p)u^{(1)\dagger}(p)-v^{(1)}(-p)v^{(1)\dagger}(-p) 
  \notag  \\ &=(\gamma^ip_i+m-g_2\omega_1^2\gamma_0\gamma_5)
    \gamma_0  (\mathbb{1}_{4} -Q)\,,
\end{align} 
\begin{align}  \label{uu2+vv2dag}
& u^{(2)}(p)u^{(2)\dagger}(p)+v^{(2)}(-p)v^{(2)\dagger}(-p) \notag \\ 
&=\omega_2  (\mathbb{1}_{4} +Q)\,,
\end{align}
\begin{align}\label{uu2-vv2dag}
& u^{(2)}(p)u^{(2)\dagger}(p)-v^{(2)}(-p)v^{(2)\dagger}(-p)  \notag \\   
&=(\gamma^ip_i+m-g_2\omega_2^2\gamma_0\gamma_5)
\gamma_0 (\mathbb{1}_{4} +Q)\,,
\end{align}
\begin{align}
&U^{(1)}(p)U^{(1)\dagger}(p)+V^{(1)}(-p)V^{(1)\dagger}(-p) 
\notag \\ &=W_1  (\mathbb{1}_{4} -Q)\,,
\end{align}  
\begin{align}
&  U^{(1)}(p)U^{(1)\dagger}(p)-V^{(1)}(-p)V^{(1)\dagger}(-p)   \notag \\ 
&=(\gamma^ip_i+m-g_2W_1^2\gamma_0\gamma_5)\gamma_0
 (\mathbb{1}_{4} -Q)\,,
\end{align}
\begin{align}
&  U^{(2)}(p)U^{(2)\dagger}(p)+V^{(2)}(-p)V^{(2)\dagger}(-p)
\notag \\ &=W_2 (\mathbb{1}_{4} +Q)\,,
\end{align} 
\begin{align}
&  U^{(2)}(p)U^{(2)\dagger}(p)-V^{(2)}(-p)V^{(2)\dagger}(-p)  \notag  \\ 
&=(\gamma^ip_i+m-g_2W_2^2\gamma_0\gamma_5)
\gamma_0 (\mathbb{1}_{4} +Q)\,.
\end{align}
\setcounter{equation}{0}
\renewcommand{\theequation}{A.\arabic{equation}}
\providecommand{\href}[2]{#2}
%.............................................................................

\end{document}